\documentclass[a4paper,flegn]{cas_dc}
\usepackage{amsmath,amsfonts}
\usepackage{amsthm}
\usepackage{array}
\usepackage[caption=false,font=normalsize,labelfont=sf,textfont=sf]{subfig}
\usepackage{textcomp}
\usepackage{stfloats}
\usepackage{verbatim}
\usepackage{graphicx}
\usepackage{nomencl}
\usepackage{etoolbox}
\usepackage{ifthen}
\usepackage{booktabs}
\usepackage[ruled]{algorithm2e}
\usepackage{appendix}
\usepackage{multirow}
\usepackage{bm}
\usepackage{makecell}
\usepackage{algpseudocode}
\allowdisplaybreaks 
\usepackage[subtle,tracking=normal]{savetrees}
\usepackage{float}
\usepackage{url}
\usepackage{hyperref} 

\usepackage[numbers,sort&compress]{natbib}
\usepackage{nomencl}
\makenomenclature
\renewcommand{\nomgroup}[1]{%
    \ifthenelse{\equal{#1}{A}}{\item[\textbf{Abbreviations}]}{%
    \ifthenelse{\equal{#1}{B}}{\item[\textbf{Sets and Metrics}]}{%
    \ifthenelse{\equal{#1}{C}}{\item[\textbf{Functions}]}{%
    \ifthenelse{\equal{#1}{D}}{\item[\textbf{Parameters}]}{
    \ifthenelse{\equal{#1}{E}}{\item[\textbf{Decision Variables}]}{}}}}}}
\usepackage{multicol}
\usepackage{framed} 
\usepackage{etoolbox} 
\usepackage{float}
\usepackage{placeins}
\usepackage{threeparttable} 

\def\tsc#1{\csdef{#1}{\textsc{\lowercase{#1}}\xspace}}
\tsc{WGM}
\tsc{QE}

\newtheoremstyle{normalremark} 
  {3pt} 
  {3pt} 
  {} 
  {} 
  {\bfseries} 
  {.} 
  {5pt plus 1pt minus 1pt} 
  {} 
\theoremstyle{normalremark}

\newtheorem{remark}{Remark}

\newtheorem{lemma}{Lemma}

\begin{document}
\let\WriteBookmarks\relax
\def\floatpagepagefraction{1}
\def\textpagefraction{.001}
\let\printorcid\relax 

\shorttitle{\rmfamily N.Qi et al. Capacity Credit Evaluation of Generalized Energy Storage}    

\shortauthors{\rmfamily N. Qi et al.}

\title[mode = title]{Capacity Credit Evaluation of Generalized Energy Storage Considering Strategic Capacity Withholding and Decision-Dependent Uncertainty}

\author[1]{Ning Qi}
\credit{Conceptualization, Modeling, Methodology, Software, Writing}

\author[2]{Pierre Pinson}
\credit{Methodology, Revision}

\author[3]{Mads R. Almassalkhi}
\credit{Methodology, Revision}

\author[4]{Yingrui Zhuang}
\credit{Revision, Software}

\author[5]{Yifan Su}
\credit{Revision}

\author[4]{Feng Liu}
\credit{Supervision, Revision, Funding Support}
\cormark[1]
\ead{lfeng@tsinghua.edu.cn}

\address[1]{Department of Earth and Environmental Engineering, Columbia University, New York, NY 10027, USA}
\address[2]{Dyson School of Design Engineering, Imperial College London, SW7 2AZ London, UK}
\address[3]{Department of Electrical and Biomedical Engineering, University of Vermont, Burlington, VT 05405, USA}
\address[4]{Department of Electrical Engineering, Tsinghua University, Beijing 100084, China}
\address[5]{State Key Laboratory of Power Transmission Equipment Technology, School of Electrical Engineering, Chongqing University, Chongqing 400044, China}

\cortext[1]{Corresponding author} 

\begin{abstract}
This paper proposes a novel capacity credit evaluation framework to accurately quantify the contribution of generalized energy storage (GES) to resource adequacy, considering both strategic capacity withholding and decision-dependent uncertainty (DDU). To this end, we establish a market-oriented risk-averse coordinated dispatch method to capture the cross-market reliable operation of GES. The proposed method is sequentially implemented along with the Monte Carlo simulation process, coordinating the pre-dispatched price arbitrage and capacity withholding in the energy market with adequacy-oriented re-dispatch during capacity market calls. In addition to decision-independent uncertainties in operational states and baseline behavior, we explicitly address the inherent DDU of GES (i.e., the uncertainty of available discharge capacity affected by the incentives and accumulated discomfort) during the re-dispatch stage using the proposed \textcolor{blue}{data-driven} distributional robust chance-constrained approach. Furthermore, a capacity credit metric called equivalent storage capacity substitution is introduced to quantify the equivalent deterministic storage capacity of uncertain GES. Simulations on the modified IEEE RTS-79 benchmark system with 20 years real-world data from Elia demonstrate that the proposed method yields accurate capacity credit and improved economic performance. We show that the capacity credit of GES increases with more strategic capacity withholding but decreases with more DDU levels. Key factors, such as capacity withholding and DDU structure impacting GES's capacity credit are analyzed with insights into capacity market decision-making.
\end{abstract}


\begin{highlights}
\item A long-term decision-dependent uncertainty model for generalized energy storage
\item A market-oriented risk-averse coordinated dispatch simulation method
\item Data-driven distributionally robust chance-constrained optimization under decision-dependent uncertainty
\item A novel capacity credit metric called equivalent storage capacity substitution
\item Numerical studies based on Elia system with ground-truth datasets spanning 20 years

\end{highlights}

\maketitle

\section{Introduction}

\subsection{Background and motivation}
As an emerging concept, generalized energy storage (GES) involves both physical energy storage (ES) and virtual energy storage (VES) offered by demand response (DR)~\cite{qi2023chance}. GES can represent an individual unit or a portfolio of heterogeneous quasi-storage units uniformly, which simplifies the integration and coordination of massive and fragmented flexible resources. GES can participate not only in energy and ancillary service markets~\cite{zou2015evaluating}, but are also expected to be major contributors to the capacity market~\cite{IEAREPORT}, where they commit to supplying capacity during system emergencies under long-term contracts. Hence, it is crucial to accurately pre-evaluate GES's contribution to resource adequacy, as this effort is vital for the long-term planning and capacity market pricing~\cite{mertens2021capacity}. 

Traditional resource adequacy evaluation is primarily based on capacity credit (CC), which quantifies the ability
of incremental renewable generation to displace conventional generation without compromising system reliability. This typically involves generating the power output time series of renewable generation via Monte Carlo simulation, which is then used to calculate CC based on various metrics, including Effective Load Carrying Capability (ELCC)~\cite{chen2014effective}, Equivalent Firm Capacity (EFC)~\cite{zachary2012probability}, Equivalent Conventional Capacity (ECC)~\cite{dent2014defining}, and Equivalent Generation Capacity Substitution (EGCS)~\cite{zhou2016}. However, it becomes challenging for the energy-limited GES, as it requires an additional simulation of the state of charge (SoC), which is highly dependent on operational strategy and various exogenous and endogenous factors.

\subsection{Literature review}

Previous works typically adopt \textit{adequacy-oriented dispatch methods} to evaluate CC of GES. Based on the difference to handling real-time failures, these methods can be broadly categorized into fixed dispatch, greedy dispatch, perfect foresight, and re-dispatch methods. Fixed dispatch is typically implemented on peak days with fixed schedule for each simulation day regardless of real-time failures. For instance, the operational strategy of ES is generated based on peak shaving optimization~\cite{zhou2016,dratsas2021battery,wang2024multi}, which aims to reduce the peak load and potential loss of load on the peak days. The CC of VES is evaluated in~\cite{safdarian,zeng2023quantifying}, where the operational strategy of VES is generated using the fixed average responsive load profiles. However, under fixed dispatch, GES operations are determined before the simulation of system states, without adjusting the GES operations in response to real-time failures. This will result in under-utilization of GES flexibility and underestimated CC values. Unlike fixed dispatch, ``greedy dispatch'' is proposed in~\cite{evans2019,edwards2017assessing}, where ES remains fully charged during normal system states and only provides capacity during emergencies. The CC of ES is evaluated based on the planning optimization framework where ES is treated as a transmission asset~\cite{sehloff2024assessing}. The perfect foresight of loss-of-load events is even assumed in~\cite{stenclik2018energy} for ES scheduling. Although greedy dispatch
\newpage
\nomenclature[A]{BES/GES/VES}{Battery/generalized/virtual energy storage}
\nomenclature[A]{CC}{Capacity credit}
\nomenclature[A]{DIU/DDU}{Decision-independent/dependent uncertainty}
\nomenclature[A]{DR}{Demand response}
\nomenclature[A]{EV}{Electric vehicle}
\nomenclature[A]{SMCS}{Sequential Monte Carlo simulation}
\nomenclature[A]{SoC}{State of charge}
\nomenclature[A]{TCL}{Thermostatically controlled load}
\nomenclature[B]{$\mathcal{B}/\mathcal{T}_{\text{N}}^{(\delta)}$}{Set for buses of network/time slots during normal states}
\nomenclature[B]{$\mathcal{T}_{\text{E}}^{(k)}/\mathcal{T}_{\text{R}}^{(k)}$}{Set for time slots during emergency states/recovery states}
\nomenclature[B]{\(\bm{x}/\bm{\xi}\)}{Set for decision variables/DDU}
\nomenclature[B]{\text{ECC/EFC}}{Equivalent conventional/firm capability}
\nomenclature[B]{\text{ELCC}}{Effective load carrying capabiliy}
\nomenclature[B]{\text{EGCS}}{Equivalent generation capacity substituted}
\nomenclature[B]{\text{ESCS}}{Equivalent storage capacity substitution}
\nomenclature[B]{$\text{EENS}^{\text{P/T}}$}{Practical/theoretical expected energy not served}
\nomenclature[C]{\(R(\cdot)\)}{Reliability function}
\nomenclature[C]{\(F^{-1}(\cdot)\)}{Inversed cumulative distribution function}
\nomenclature[C]{\(g(\cdot)/h(\cdot)\)}{DDU function of incentive/discomfort effect}
\nomenclature[C]{\(\mathbb{P(\cdot)}\)}{Probability function of chance constraint}
\nomenclature[D]{\(\overline{SoC}_{i\text{,}t}/ \underline{SoC}_{i\text{,}t}\)}{Upper/lower SoC bounds of GES unit at bus \textit{i}, time slot \textit{t}}
\nomenclature[D]{\(\overline{P}_{i\text{,}t}/ \underline{P}_{i\text{,}t}\)}{Upper/lower power bounds of GES unit at bus \textit{i}, time slot \textit{t}}
\nomenclature[D]{\(SoC_{i,t}^{\text{B}}\)}{Baseline SoC of GES unit at bus \textit{i}, time slot \textit{t}}
\nomenclature[D]{\(\varphi_{i}/NC_{t}\)}{Capacity
allocation ratio of GES unit at bus \textit{i}/system net capacity at time slot \textit{t}}
\nomenclature[D]{\(\eta_{\text{c/d}\text{,}i}/\varepsilon_{i}/E_{i}\)}{Charge/discharge efficiency/self-discharge rate/energy capacity of GES at bus \textit{i}}
\nomenclature[D]{\(\Delta t/\textit{T}\)}{Time-step/duration of dispatch}
\nomenclature[D]{\(\mu/\sigma\)}{Mean/standard deviation of the distribution}
\nomenclature[D]{\(c^{\text{CM}}/c_{t}^{\text{EM}}\)}{Capacity/energy market price at time slot \textit{t}}
\nomenclature[D]{\(\alpha_{i}/\beta_{i}\)}{ Incentive elasticity factor/discomfort-aversion factors of SoC bound of GES unit at bus \textit{i}}
\nomenclature[D]{\(\rho/\lambda\)}{ Weight coefficient between discomfort from past calls and current one/between response intensity and SoC-based discomfort}
\nomenclature[D]{\({P}_{i\text{,}t}^{\text{CG/RG}}/{P}_{i\text{,}t}^{\text{L}}\)}{Available power of conventional generator/ renewable generator/load at bus \textit{i}, time slot \textit{t}}
\nomenclature[D]{\(\epsilon/\text{CoV}\)}{Confidence level of chance constraint/convergence criterion of SMCS}
\nomenclature[E]{\(P_{\text{c/d}\text{,}i\text{,}t}^{\text{PD}}/P_{\text{c/d}\text{,}i\text{,}t}^{\text{RD}}\)}{Pre-dispatched/re-dispatched charge/ discharge power of GES unit at bus \textit{i}, time slot \textit{t}}
\nomenclature[E]{\(P_{i\text{,}t}^{\text{LC}}\)}{Curtailed load power at bus \textit{i}, time slot \textit{t}}
\nomenclature[E]{\(SoC_{i\text{,}t}^{\text{PD/RD}}\)}{Pre-dispatched/re-dispatched SoC of GES unit at bus \textit{i}, time slot \textit{t}}
\nomenclature[E]{\(D_{i\text{,}t}\)}{Response discomfort of GES unit at bus \textit{i}, time slot \textit{t}}
\nomenclature[E]{\(S_{\text{c/d}\text{,}i\text{,}t}^{\text{PD}}/S_{\text{c/d}\text{,}i\text{,}t}^{\text{RD}}\)}{Pre-dispatched /re-dispatched charge/ discharge states of GES unit at bus \textit{i}, time slot \textit{t}}
\setlength{\nomlabelwidth}{2.2cm}
\setlength{\nomitemsep}{-\parsep}
\twocolumn[{%
    \begin{center}
    \end{center}
    \begin{framed}
    \begin{multicols}{2}
    \printnomenclature
    \end{multicols}
    \end{framed}
}]

\noindent and perfect foresight manage to adjust the GES operational profile during real-time failures, they overlook the potential of GES to provide other services (e.g., price arbitrage and frequency regulation). These two methods have been pointed out by ENTSO-E~\cite{ENTSO} as optimistic and unrealistic, which tend to overestimate the CC of GES. Given the limitations of the aforementioned methods, a recent work~\cite{dratsas2023real} proposes a real-time re-dispatch method where ES follows the day-ahead unit commitment strategy but switches to adequacy-oriented dispatch during real-time failures. However, the planned schedule aims to minimize system costs, making it suitable only for centralized (non-merchant) storage. It is not applicable to decentralized (merchant) GES resources, which tend to prioritize profit maximization~\cite{qi2023portfolio,xiao2023integrated} and withhold capacity to capture future opportunities~\cite{zhou2024energy,ma2025comparative}. Instead, developing a market-oriented dispatch method can better capture the realistic operations of merchant GES resources. A stochastic dynamic programming approach is proposed in~\cite{kim2021stochastic} to co-optimize ES for providing energy-shifting and resource adequacy services simultaneously, explicitly accounting for the uncertainty of loss-of-load events in the decision-making. The stochastic framework has been extended to consider the additional provision of frequency regulations in~\cite{kim2024assessing}. The strategy for uninterruptible power supply storage is generated using a sequential decision-making approach to maximize rewards from both price arbitrage and capacity provision~\cite{yong2023capacity}. However, the inherent limitation of the existing dispatch methods is the treatment of the uncertainties of GES, as they rely on deterministic optimization or only address the uncertainty of loss of load events, which fails to mitigate the risks from GES and poses potential risks to system adequacy.

In addition to the dispatch method, the CC of GES can also be affected by various factors, such as power and energy capacity~\cite{zhou2016,dratsas2021battery,kim2021stochastic,konstantelos2018capacity,mertens2021capacity,dratsas2023real,parks2019declining,yong2023capacity}, efficiency~\cite{zhou2016,dratsas2021battery,konstantelos2018capacity}, network constraints~\cite{wang2021crediting,liu2020assessment}, generation mix~\cite{denholm2020potential,wang2021crediting,dratsas2021battery}, etc. However, a key endogenous factor that has been underexplored is the inherent \textit{uncertainty} of GES. Some of the existing works overlook the uncertainty of GES~\cite{liu2020assessment,mertens2021capacity,wang2021crediting,parks2019declining,zhou2016,evans2019,edwards2017assessing,stenclik2018energy,kim2021stochastic,safdarian,denholm2020potential}. Nevertheless, GES availability is practically subject to reliability attribute~\cite{zhang2024prediction}, baseline behavior~\cite{qi2022reliability}, etc., which can significantly reduce its contribution to resource adequacy. The forced outage is accounted for ES availability using a two-state Markov chain~\cite{dratsas2023real,dratsas2021battery,konstantelos2018capacity}. The availability of VES is primarily impacted by incentive and occupant discomfort, Markov chains~\cite{kwag2014} and probabilistic distributions~\cite{nikzad2014} are typically used to simulate its operational state and power output. However, most of the uncertainty mentioned above is the decision-dependent uncertainty (DIU), which is independent of the decision-making process. While, some stochasticity is inherently decision-dependent uncertainty (DDU), where the uncertainty realization will be affected by decisions. This type of uncertainty is often overlooked or simplified as DIU with static and known distributions. For instance, battery degradation can be modeled as a function of discharge cycles, depth of discharge, etc.~\cite{cheng2021operational}, making the available capacity inherently dependent on its dispatch. The magnitude and duration of DR can impact the probability of manual overrides for thermostatically controlled loads (TCL), leading to a decision-dependent reduction in response capacity~\cite{kane2020data}. Future price forecasts have been shown to alter the distribution of price-responsive demand~\cite{ellinas2024decision}, thereby affecting the availability of DR. Recent works~\cite{qi2023chance,pan2022modeling,hu2021decision} have recognized the importance of DDU modeling and have proposed tractable optimization methods to mitigate the risks associated with DDU. However, these works primarily address short-term DDU while overlooking long-term decision dependency. A pioneering work~\cite{zeng} proposes a long-term DDU model for DR in CC evaluation, but the model is only used to calculate the consequences of overlooking DDU and is not incorporated into the dispatch method. To the best of our knowledge, no research has yet developed a market-oriented risk-averse dispatch method that explicitly incorporates the long-term DDU of GES in the CC evaluation. 

While the aforementioned studies provide valuable theoretical insights, it is equally important to understand real-world practices in capacity markets. ES and DR have participated in Reliability Pricing Model or Forward Capacity Market across the United States, Europe and Asian countries~\cite{khan2018demand,mccullough2020exactly,lu2018reliability}.  However, consensus has not been reached on how to quantify the CC of GES. For instance, NYISO suggests a derating CC for ES based on the discharge duration. PJM and CASIO recommend simulation approaches using ELCC metric with ES dispatch and the historical performance of DR. Furthermore, CAISO faces significant challenges with the response unavailability of GES and highlights that even after accounting for random failures, around 15\% response unavailability still occurs during peak events~\cite{CAISO}, resulting in substantial costs for reserves and load curtailment. Therefore, developing a more accurate availability model and realistic dispatch method is crucial for CC evaluation of GES, especially given the demand for long-duration and frequent capacity supply in future renewable-dominated power systems.

\subsection{Research gap}

We summarize the existing literature on CC evaluation of GES and compare these methods with the evaluation framework proposed in this paper, as shown in Table~\ref{literature review}. The primary research gaps that remain inadequately addressed in the current literature are outlined as follows:

(1) CC evaluation of GES typically adopts adequacy-oriented dispatch (e.g., fixed dispatch~\cite{zhou2016,dratsas2021battery,wang2024multi,konstantelos2018capacity,denholm2020potential,wang2021crediting,liu2020assessment,parks2019declining,safdarian,zeng2023quantifying,kwag2014,nikzad2014}, greedy dispatch~\cite{evans2019,edwards2017assessing,sehloff2024assessing,stenclik2018energy}, coordinated dispatch~\cite{dratsas2023real}) to assess its contribution during emergencies. However, adequacy-oriented dispatch is limited to centralized or non-merchant storage and fails to account for the market behavior of GES. Existing market-oriented dispatch works~\cite{yong2023capacity,kim2021stochastic,kim2024assessing} only account for the cross-market behavior by coordinating multiple services, but overlook the economic capacity withholding behavior~\cite{zhou2024energy,ma2025comparative}. Therefore, to enhance the accuracy of CC evaluation, it is crucial to develop a market-oriented dispatch framework to account for both cross-market and economic capacity withholding behavior of GES. 

(2) Most existing studies consider only deterministic GES~\cite{liu2020assessment,mertens2021capacity,wang2021crediting,parks2019declining,zhou2016,evans2019,edwards2017assessing,stenclik2018energy,kim2021stochastic,safdarian,denholm2020potential} or GES with DIUs~\cite{dratsas2023real,dratsas2021battery,konstantelos2018capacity,kwag2014,nikzad2014} in the CC evaluation, overlooking its inherent DDUs. This omission fails to capture the decision-dependent characteristics of GES, leading to inaccurate CC evaluation. Although studies such as~\cite{qi2022reliability} and~\cite{zeng} account for DDUs, they do not model long-term DDUs or explicitly incorporate them into the dispatch method. Therefore, designing a risk-averse dispatch method to address the long-term DDU in the CC evaluation is necessary.

(3) Existing CC evaluation usually focus only on theoretical performance calculations without explicitly evaluating the practical consequences induced by DDU~\cite{qi2022reliability,zeng}. In addition, CC metrics, e.g., EFC~\cite{zeng}, EGCS~\cite{zhou2016} are proposed to represent how much generation or load capacity can be substituted by GES. However, most GES resources operate more like ES by charging and discharging to contribute to resource adequacy. Thus, to maintain the inter-temporal information and address long-duration power shortages, it is more effective to replace the stochastic GES with an equivalent deterministic ES and incorporate it into the co-planning of the transmission system and storage~\cite{gan2019security}.

\begin{table*}[!ht]
\footnotesize\rmfamily
  \centering
  \begin{threeparttable}
  \caption{\rmfamily Comparison of this paper with other related literature.}
  \setlength{\tabcolsep}{0mm}{
      \begin{tabular}{c c c c c c c c c}
    \toprule
    \multicolumn{1}{c}{\multirow{2}[2]{*}{Reference}} & \multicolumn{2}{c}{Dispatch Framework} & \multicolumn{2}{c}{Uncertainty Consideration} & \multicolumn{3}{c}{Evaluation Framework} \\
          & Adequacy-Oriented & Market-Oriented & DIU & DDU & Object & Metric & Performance \\
    \midrule
\cite{zhou2016,dratsas2021battery,wang2024multi,konstantelos2018capacity,denholm2020potential,wang2021crediting,liu2020assessment,parks2019declining}     &  $\checkmark$ (Fixed Dispatch)    &   $\times$    &   $\checkmark$ (MCS)    &    $\times$   &  ES     &  ECC/EGCS/EFC/ELCC     & Theoretical \\
\cite{safdarian,zeng2023quantifying,kwag2014,nikzad2014}        &  $\checkmark$ (Fixed Dispatch)    &  $\times$     &   $\checkmark$ (MCS)    &   $\times$     &   VES    &   Reliability Indices/ECC    & Theoretical \\
\cite{evans2019,edwards2017assessing,sehloff2024assessing,stenclik2018energy}       &  $\checkmark$ (Greedy Dispatch)    &   $\times$    &   $\checkmark$ (MCS)    &   $\times$    &   ES    &   ECC/EGCS/EFC/ELCC    &  Theoretical\\
\cite{dratsas2023real}      &  $\checkmark$ (Coordinated Dispatch)   &     $\times$    &   $\checkmark$ (MCS)    &   $\times$    &   ES    &   EFC    &  Theoretical\\
\cite{yong2023capacity}      &  $\times$   &    $\checkmark$ (Coordinated Dispatch)    &   $\checkmark$ (MCS)    &   $\times$    &   ES    &   ELCC    &  Theoretical\\
\cite{kim2021stochastic,kim2024assessing}      &  $\times$   &    $\checkmark$ (Coordinated Dispatch)    &   $\checkmark$ (MCS+SO)    &   $\times$    &   ES    &   ELCC    &  Theoretical\\
\cite{qi2022reliability}      &  $\checkmark$ (Fixed Dispatch)  &    $\times$     &   $\checkmark$ (MCS)    &   $\checkmark$ (CCO)    &   VES    &   Reliability Indices    &  Theoretical+Pratical\\
\cite{zeng}      &  $\checkmark$ (Fixed Dispatch)  &    $\times$     &   $\checkmark$ (MCS)    &   $\checkmark$ (MCS)    &   VES    &   ECC    &  Theoretical+Pratical\\
This Paper      & $\times$    &    $\checkmark$ (Coordinated Dispatch+ECW)    &   $\checkmark$ (MCS)    &   $\checkmark$ (CCO)    &  ES+VES    &   ECC/EGCS    &  Theoretical+Pratical\\
    \bottomrule
    \end{tabular}%
    }\label{literature review}
    \begin{tablenotes}
\item[a] MCS: Monte Carlo simulation, SO: stochastic optimization, CCO: chance-constrained optimization, ECW: Economic capacity withholding.
\end{tablenotes}
\end{threeparttable}\vspace{-0.5cm}
\end{table*}%

\subsection{Contributions}

Motivated by this background, this paper proposes a novel CC evaluation framework for GES, which concurrently incorporates the market-oriented coordinated dispatch and risk management of DDU. The main contributions are fourfold:

\textbf{(1) Modeling:} A novel availability model of GES is established for CC evaluation. In addition to modeling DIUs in operational states and baseline power and SoC bounds, we construct a convex long-term DDU model to capture the available discharge capacity affected by capacity market incentive and accumulated discomfort during capacity market calls. The proposed DDU model is built on our recent work~\cite{qi2023chance} and incorporates discomfort ``memory'' from previous capacity market calls. Compared to the deterministic model~\cite{zhou2016}, the DIU model~\cite{kwag2014}, and the non-convex DDU model~\cite{zeng} in the literature, our model can better capture the decision-dependent characteristics of GES availability and is more easily integrated into dispatch optimization.

\textbf{(2) Dispatch Framework}: We propose a market-oriented risk-averse coordinated dispatch method to simulate cross-market strategic behaviors of GES and mitigate DDU-induced risk in CC evaluation. Compared to adequacy-oriented dispatch methods~\cite{zhou2016,evans2019} that address GES operations during emergencies only, our method coordinates price arbitrage in the energy market with adequacy-oriented re-dispatch during capacity market calls. Compared to existing market-oriented dispatch methods~\cite{yong2023capacity, dratsas2023real}, our method can also \textcolor{blue}{account for the strategic} SoC withholding behaviors in the pre-dispatch to reserve the available capacity for capacity provision, as well as address the inherent DDU of GES in the re-dispatch via distributionally chance-constrained optimization.

\textbf{(3) Solution Methodology}: We propose a tractable reformulation method for the distributionally robust chance-constrained optimization under DDU. Specifically, we introduce a data-driven reformulation approach based on available data and information on DDU to approximate and substitute the DDU terms within the optimization model. Additionally, we provide a theoretical proof demonstrating the convergence of the proposed method. Compared to previous works~\cite{qi2023chance,qi2023portfolio,pan2022modeling} that assume either ambiguous or complete knowledge of uncertainty distributions, our proposed method effectively addresses data scarcity issues in practice, while simultaneously avoiding the overly conservative decisions associated with robust optimization methods~\cite{su2022multi}.

\textbf{(4) Evaluation Framework}: The practical performance considering the DDU consequence is further addressed in the CC evaluation framework. Simulations show the importance of incorporating cross-market strategic behavior and DDU in CC evaluation, resulting in a more accurate CC value and improved economic performance. Moreover, we introduce a novel CC metric, equivalent storage capacity substitution (ESCS), to quantify the equivalent deterministic storage capacity of uncertain GES. This metric preserves the inter-temporal properties of GES and can be used for the co-planning of transmission systems and GES. Additionally, key factors (e.g., capacity withholding ratio, DDU structure) impacting the CC of GES are analyzed with helpful insights into capacity market decision-making.

\subsection{Paper Organization}

We organize the remainder of the paper as follows. Section~\ref{dispatch} presents the formulation, solution methodology and pratical simulation of the market-oriented risk-averse coordinated dispatch method for GES. Section~\ref{evaluation} introduces the novel CC evaluation methodology of GES, which integrates the proposed coordinated dispatch method.  Numerical studies and sensitivity analysis are proposed in Section~\ref{case} and Section~\ref{SA} to demonstrate the effectiveness of the proposed method. Finally, section~\ref{conclusion} concludes the paper.

\section{Market-Oriented Risk-Averse Coordinated Dispatch}\label{dispatch}

\subsection{Motivation}\label{motivation}
In liberalized markets, GES resources (including battery, TCL, electric vehicle (EV), etc.) can actively participate in both short-term energy markets and the long-term capacity market. In energy markets, GES aims for price arbitrage by shifting energy in response to short-term prices. Meanwhile, the capacity market operates separately as a remuneration mechanism designed to ensure sufficient capacity is available to meet future demand. The capacity market sets obligations for participants, e.g., the duration and frequency of response. GES participants are remunerated based on their practical performance, with additional rewards for excess performance or penalties for non-performance. Specifically, VES resources can engage in the price-based and emergency DR programs to participate in energy market and capacity market~\cite{PJMDR}. 

\begin{figure}[htp]
\setlength{\abovecaptionskip}{-0.1cm}  
\setlength{\belowcaptionskip}{-0.1cm} 
  \begin{center}  \includegraphics[width=1\columnwidth]{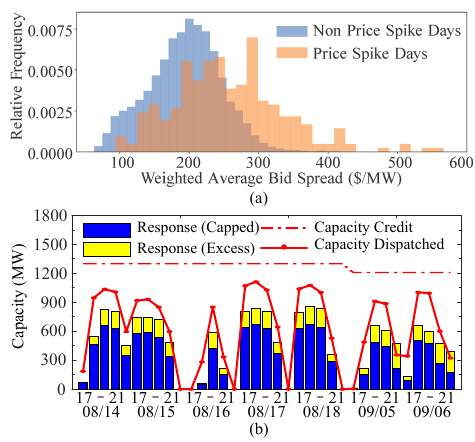}
    \caption{Practical performance of GES in CAISO (a) strategic capacity withholding in energy market and (b) capacity availability in capacity market.}\label{motivate}
  \end{center}\vspace{-0.5cm}
\end{figure}

Considering that GES optimally allocates its capacity across multiple services to maximize profit~\cite{qi2023portfolio,xiao2023integrated}, it is essential to understand its cross-market strategic behavior. As demonstrated by practical performance in CAISO, GES may strategically withhold capacity by raising bid prices on normal days to capture potential opportunities during price-spike days, as shown in Fig.~\ref{motivate} (a). It can be reasonably inferred that GES will also strategically preserve SoC in the energy markets to secure potentially higher-profit opportunities in the capacity market, while simultaneously avoiding capacity market penalties. Such withholding behavior reduces its bidding volume and corresponding revenues from the energy market. On the other hand, active participation in energy markets consumes stored energy and thus reduces the available capacity for emergency situations, potentially lowering its overall CC. Furthermore, as illustrated in Fig.~\ref{motivate}(b), approximately one-third of VES unavailability was observed during 2020–2021, although this unavailability significantly decreased to around 15\% during 2023–2024. However, the practical CC of VES appears to remain lower than the theoretical values, showing dynamic characteristics and decreasing further with longer response durations. In fact, the available capacity of GES depends on both physical capacity (e.g., capacity degradation) and response willingness (e.g., occupant discomfort), which is inherently impacted by incentives and previous dispatch decisions. This stochasticity, known as DDU, can potentially lead to increased response unavailability if not well-addressed. Therefore, it is necessary to employ a risk-averse approach to effectively hedge against the risks arising from DDU. We next present problem formulation and solution methodology for market-oriented risk-averse coordinated dispatch, which simulates the realistic cross-market operations of GES and mitigate the risk associated with DDU.

\subsection{Problem Formulation}\label{dispatchformulation}

\begin{figure}[htp]
\setlength{\abovecaptionskip}{-0.1cm}  
\setlength{\belowcaptionskip}{-0.1cm} 
  \begin{center}  \includegraphics[width=1\columnwidth]{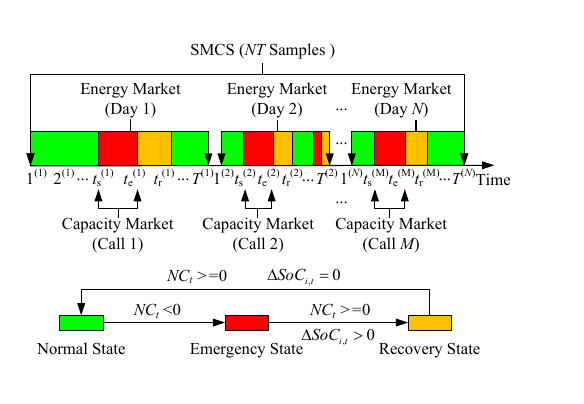}
    \caption{Diagram of market-oriented risk-averse coordinated dispatch for GES.}\label{Vdispatch}
  \end{center}\vspace{-0.5cm}
\end{figure}

The market-driven risk-averse coordinated dispatch is sequentially implemented within 
SMCS over  \(N\) days, with each day consisting of \(T\) samples. \(T\) represents the duration of the optimization horizon for each day. The dispatch of each day may involve transitions between three distinct system states: normal state (green color), emergency state (red color), and recovery state (orange color), as depicted in Fig.~\ref{Vdispatch}. These transitions are triggered by the system net capacity and deviations of SoC in pre-dispatch and re-dispatch, which are defined in equations~\eqref{defRC} and~\eqref{defdelta}, respectively. 
$P_{i,t}^{\text{CG}}$ and $P_{i,t}^{\text{RG,}}$ denote the available capacity of conventional and renewable generation, respectively. $P_{i,t}^{\text{LD}}$ denotes Load power. 
$SoC_{i,t}^{\text{PD}}$ and $SoC_{i,t}^{\text{RD}}$ denote the SoC in pre-dispatch and re-dispatch, respectively.
\begin{subequations}~\label{defSMCS}
\begin{align}
& NC_{t}=\sum\nolimits_{i}(P_{i,t}^{\text{CG}}+P_{i,t}^{\text{RG}}-P_{i,t}^{\text{LD}})\label{defRC}\\
& \Delta SoC_{i,t}=|SoC_{i,t}^{\text{RD}}- SoC_{i,t}^{\text{PD}}|\label{defdelta}
\end{align}
\end{subequations}

We follow the following principles to generate the operational strategy of GES:
\begin{enumerate}
    \item First, we generate the pre-dispatch strategy for GES aimed at daily price arbitrage in the energy market.
    \item Normal state: when the system net capacity is sufficient and no SoC deviation is observed, i.e., $NC_{t}\geq0$ \& $\Delta SoC_{i,t}=0$, GES will follow the pre-dispatch strategy.
    \item Emergency state: when capacity deficiency events happen, i.e., $NC_{t}<0$, capacity market issues calls\footnote{``call'' refers to an action initiated by the system operator to request or signal capacity market participants to deliver their contracted capacity.} and GES will be re-dispatched to reduce the loss of load.
    \item Recovery state: when capacity deficiency event ends and SoC deviation is observed, i.e., $RC_{t}\geq0$ \& $\Delta SoC_{i,t}>0$, GES will recover its SoC to the pre-dispatch level.
\end{enumerate}

To facilitate the integration and coordination of massive and fragmented flexible resources, we leverage the GES model~\cite{qi2023chance} for the following problem formulation. The proposed model is designed as a unified formulation, applicable to a general class of storage-like resources. Such a unified representation allows the model to flexibly describe both individual units and aggregations of diverse GES portfolios, including but not limited to batteries, TCLs, and EVs. In addition to the GES model, the physical capacity degradation of GES is explicitly considered by adopting the linear degradation model proposed in~\cite{jiang2019unified}, and accordingly, the available SoC bounds are updated each time slot.

\textbf{(i) Normal State:} The price-taker price arbitrage in the energy market for GES at each bus $i\in \mathcal{B}$ is formulated in~\eqref{normalstate}. The objective function~\eqref{objectDA}~maximizes the energy market profit for GES. Constraints~\eqref{soc-threshold} are the newly introduced constraints to limit the minimum SoC withholding. Constraints~\eqref{soc-power}~define the relationship between charge/discharge actions and SoC. Constraints~\eqref{soc-bound}~limit the upper and lower bounds on SoC. Constraints~\eqref{soc-balance} ensure that the SoC at the final time slot equals that of the initial time slot. Constraints~\eqref{power-bound1}-\eqref{power-bound2}~limit the upper and lower charge/discharge power. Constraints~\eqref{scd} prevent simultaneous charging and discharging at the same time slot. 
\begin{subequations}~\label{normalstate}
\begin{align}
& \max \ \sum\nolimits_{t\in \mathcal{T}_{\text{N}}^{(\delta)}}{c_{t}^{\text{EM}}(P_{\text{d}\text{,}i\text{,}t}^{\text{PD}}-P_{\text{c}\text{,}i\text{,}t}^{\text{PD}})}\label{objectDA}\\\hspace{-0.25cm}
\text{s.t.}\hspace{0.25cm} & \underline{SoC}^{\text{CW}}\le SoC_{i\text{,}t}^{\text{PD}}\text{, }\forall t\in \mathcal{T}_{\text{N}}^{(\delta)}\label{soc-threshold}\\
&SoC_{i\text{,}t+1}^{\text{PD}}=(\eta _{\text{c}\text{,}i}P_{\text{c}\text{,}i\text{,}t}^{\text{PD}}-P_{\text{d}\text{,}i\text{,}t}^{\text{PD}}/\eta _{\text{d}\text{,}i})\Delta t/E_{i}\label{soc-power}\\
& \hspace{1.34cm}+(1-\varepsilon_{i} \Delta t)SoC_{i\text{,}t}^{\text{PD}} \text{, }\forall t\in \mathcal{T}_{\text{N}}^{(\delta)}\nonumber\\ 
 & \underline{SoC}_{i\text{,}t}\le SoC_{i\text{,}t}^{\text{PD}}\le \overline{SoC}_{i\text{,}t }\text{, }\forall t\in \mathcal{T}_{\text{N}}^{(\delta)}\label{soc-bound}\\
 & SoC_{i\text{,}T}^{\text{PD}}=SoC_{i\text{,}0}^{\text{PD}}\text{, }\forall t\in \mathcal{T}_{\text{N}}^{(\delta)}\label{soc-balance}\\
 & 0\le P_{\text{c}\text{,}i\text{,}t}^{\text{PD}}\le S_{\text{c}\text{,}i\text{,}t}\overline{P}_{\text{c}\text{,}i\text{,}t}\text{, }\forall t\in \mathcal{T}_{\text{N}}^{(\delta)}\label{power-bound1}\\
 &  0\le P_{\text{d}\text{,}i\text{,}t}^{\text{PD}}\le S_{\text{d}\text{,}i\text{,}t}\overline{P}_{\text{d}\text{,}i\text{,}t}\text{, }\forall t\in \mathcal{T}_{\text{N}}^{(\delta)}\label{power-bound2}\\
  &  0\le S_{\text{c}\text{,}i\text{,}t}^{\text{PD}}+S_{\text{d}\text{,}i\text{,}t}^{\text{PD}}\le 1\text{, }\forall t\in \mathcal{T}_{\text{N}}^{(\delta)}\label{scd}
\end{align}
\end{subequations}
\noindent where subscript $i$ and $t$ represent the bus number and time slot, respectively. Sets for time slots each day and buses are defined as $\mathcal{T}_{\text{N}}^{(\delta)}=\{1^{(\delta)}\text{,}2^{(\delta)}\text{,}\cdots\text{,}T^{(\delta)}\}\text{,}\  \delta=1\text{,}2\text{,}\cdots\text{,}N$ and $\mathcal{B}$. The superscript $\delta$ denotes the day number. Continuous decision variables for charge power, discharge power, and SoC in pre-dispatch denote $P_{\text{c}\text{,}i\text{,}t}^{\text{PD}}$, $P_{\text{d}\text{,}i\text{,}t}^{\text{PD}}$ and $SoC_{i\text{,}t}^{\text{PD}}$. Discrete decision variables for charge and discharge states in pre-dispatch denote $S_{\text{c/d}\text{,}i\text{,}t}^{\text{PD}}$. Parameters are defined as follows: $c_{t}^{\text{EM}}$ denotes the energy market price. $\overline{P}_{\text{c/d}\text{,}i\text{,}t}$ denote the charge and discharge power bounds. $\overline{SoC}_{i\text{,}t}$ and $\underline{SoC}_{i\text{,}t}$ denote the upper and lower SoC bounds. $\underline{SoC}^{\text{CW}}$ denotes the SoC withholding threshold. $\eta _{\text{c}\text{,}i}$ and $\eta _{\text{d}\text{,}i}$ denote the charge and discharge efficiency. $\varepsilon_{i}$ and $E_{i}$ denote the self-discharge rate and energy capacity.
\begin{remark}[Pre-dispatch Strategy]
GES can provide multiple short-term services to maximize their profits, which may reduce the available capacity for the capacity market. Moreover, accurately predicting all potential short-term market participation of GES remains a challenge. To simplify short-term market simulations, we introduce the constraint~\eqref{soc-threshold} to regulate the SoC threshold for capacity market participation compared to the market-oriented dispatch in~\cite{yong2023capacity}, ensuring robust CC evaluation results. It also aligns with the real-world SoC withholding behavior~\cite{ma2025comparative}. We will further investigate the impact of SoC withholding on CC in Section~\ref{SoCWithholding}. Furthermore, the price-taker price arbitrage can better capture the strategic behavior of most small-scale decentralized (merchant) GES, while the unit commitment without storage bidding strategy used in~\cite{dratsas2023real} is only appropriate for centralized (non-merchant) GES. When GES capacity becomes sufficiently large, a bidding-then-clearing framework~\cite{qi2025locational} should be adopted to generate the pre-dispatch strategy.
\end{remark}

\textbf{(ii) Emergency State:} The adequacy-oriented re-dispatch is formulated as a multi-period chance-constrained optimization in~\eqref{emergencystate}. The objective function~\eqref{objective1} aims to minimize the loss of load over the entire period of each capacity market call and across all buses. Constraint~\eqref{PB} defines the system power balance.  Curtailed load power is limited by the load power at each bus as~\eqref{LC}. Constraints~\eqref{power-bound1R}-\eqref{power-bound2R}~limit the upper and lower charge/discharge power. Constraints~\eqref{scdR} prevent simultaneous charging and discharging at the same time slot. Constraints~\eqref{soc-powerR}~define the relationship between charge/discharge actions and SoC. Constraints~\eqref{soc0} define that SoC remains consistent with the pre-dispatch at the start time $t_{\text{s}}^{(k)}$. 

\begin{figure}[!t]
      \setlength{\abovecaptionskip}{-0.1cm}  
    \setlength{\belowcaptionskip}{-0.1cm} 
  \begin{center}  \includegraphics[width=1\columnwidth]{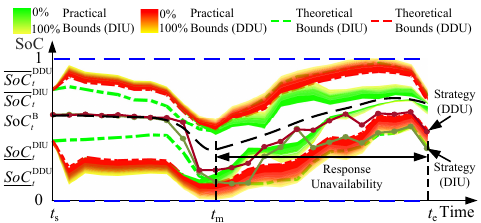}
    \caption{Comparision of SoC bounds and GES strategy under DIU and DDU.}\label{DDUbounds}
  \end{center}\vspace{-0.5cm}
\end{figure}

The available capacity of GES is an important uncertainty to be considered. During normal states, the SoC bounds of VES are endogenously determined by the occupant preferences and heterogeneity (e.g, setpoint temperature range for TCL and minimum SoC threshold for EV). Consequently, the uncertainty in SoC bounds is regarded as baseline behavior, referred to as DIU, and marked with green dash lines in Fig.~\ref{DDUbounds}. However, during emergency states, decisions and incentives from capacity market will alter this baseline behavior and affect the distribution of SoC bounds, resulting in a trade-off between inconvenience costs (i.e., discomfort during capacity market calls) and expected earnings (i.e., payment from capacity market). Hence, the SoC bounds become a DDU and are marked as red dash lines. Compared to the bounds under DIU, they will either expand up to $[0\text{,}1]$ as blue dashed lines, representing full available capacity or contract down to the average baseline performance as the black dash line, representing zero available capacity. We only provide DDU modeling for lower SoC bound as we focus on the discharge capacity. The SoC is limited by the DDU bounds by the chance-constraints~\eqref{DDU}, which ensures that the reliability of GES response falls within the confidence level (1-$\epsilon$), thereby ensuring the accuracy of CC evaluation. Constraints~\eqref{DIU-DDUup}~capture the two opposing effects of incentive and discomfort, which directly cause the SoC bounds to shift from DIU to DDU. Constraints~\eqref{spg} define the incentive effect function $g$ of capacity market price $c^{\text{CM}}$ and follows a distribution $\mathcal{G}$. Constraints~\eqref{sph} define the discomfort function of ${D}_{i,t}$ and follows a distribution $\mathcal{H}$. Constraints~\eqref{RDIN} define the mean of two distributions associated with incentive elasticity factor $\alpha_{i}$ and discomfort-aversion factor $\beta_{i}$. The long-term response discomfort with a convex form is defined as~\eqref{RDfunction}. Compared to our recent work~\cite{qi2023chance}, we further incorporate the discomfort ``memory'' in the first term to represent the response frequency and capture the accumulated discomfort prior to the $k^{th}$ call. The second term reflects the response intensity, which impacts disutility (e.g., linearized capacity degradation of ES). The final term describes the absolute SoC deviation from the average baseline performance.
\begin{subequations}\label{emergencystate}
\begin{align}
&\hspace{-0.2cm} \min\  \sum\nolimits_{t\in \mathcal{T}_{\text{E}}^{(k)}} \sum\nolimits_{i\in \mathcal{B}}{P_{i\text{,}t}^{\text{LC}}}\label{objective1}\\
&\hspace{-0.2cm}\text{s.t.}\hspace{0.1cm} NC_{i\text{,}t}+P_{i\text{,}t}^{\text{LC}}+{{P}_{\text{d}\text{,}i\text{,}t}^{\text{RD}}}-{{P}_{\text{c}\text{,}i\text{,}t}^{\text{RD}}}=0\text{, }\forall t\in \mathcal{T}_{\text{E}}^{(k)}\text{, }\forall i\in \mathcal{B}\label{PB}\\
&\hspace{-0.2cm} 0\le P_{i\text{,}t}^{\text{LC}}\le P_{i\text{,}t}^{\text{LD}}\text{, }\forall t\in \mathcal{T}_{\text{E}}^{(k)}\text{, }\forall i\in \mathcal{B}\label{LC}\\
 &\hspace{-0.2cm} 0\le P_{\text{c}\text{,}i\text{,}t}^{\text{RD}}\le S_{\text{c}\text{,}i\text{,}t}^{\text{RD}}\overline{P}_{\text{c}\text{,}i\text{,}t}\text{, }\forall t\in \mathcal{T}_{\text{E}}^{(k)}\text{, }\forall i\in \mathcal{B}\label{power-bound1R}\\
 &\hspace{-0.2cm}  0\le P_{\text{d}\text{,}i\text{,}t}^{\text{RD}}\le S_{\text{d}\text{,}i\text{,}t}^{\text{RD}}\overline{P}_{\text{d}\text{,}i\text{,}t}\text{, }\forall t\in \mathcal{T}_{\text{E}}^{(k)}\text{, }\forall i\in \mathcal{B}\label{power-bound2R}\\
  &\hspace{-0.2cm}  0\le S_{\text{c}\text{,}i\text{,}t}^{\text{RD}}+S_{\text{d}\text{,}i\text{,}t}^{\text{RD}}\le 1\text{, }\forall t\in \mathcal{T}_{\text{E}}^{(k)}\text{, }\forall i\in \mathcal{B}\label{scdR}\\
&\hspace{-0.2cm}SoC_{i\text{,}t+1}^{\text{RD}}=(\eta _{\text{c}\text{,}i}P_{\text{c}\text{,}i\text{,}t}^{\text{RD}}-P_{\text{d}\text{,}i\text{,}t}^{\text{RD}}/\eta _{\text{d}\text{,}i})\Delta t/E_{i}\label{soc-powerR}\\
&\hspace{-0.2cm} \hspace{1.34cm}+(1-\varepsilon_{i} \Delta t)SoC_{i\text{,}t}^{\text{RD}} \text{, }\forall t\in \mathcal{T}_{\text{E}}^{(k)}\text{, }\forall i\in \mathcal{B}\nonumber\\
&\hspace{-0.2cm} SoC_{i\text{,}t}^{\text{RD}}=SoC_{i\text{,}t}^{\text{PD}}\text{, } t={t}_{0}^{(k)}\text{, }\forall i\in \mathcal{B}\label{soc0}\\
 &\hspace{-0.2cm} \mathbb{P}(\underline{SoC}_{i\text{,}t}^{\text{DDU}}\le SoC_{i\text{,}t}^{\text{RD}}\le \overline{SoC}_{i\text{,}t }^{\text{DIU}})\ge 1-\epsilon \text{, }\forall t\in \mathcal{T}_{\text{E}}^{(k)}\text{, }\forall i\in \mathcal{B}\label{DDU} \\ 
&\hspace{-0.2cm}\underline{SoC}_{i\text{,}t}^{\text{DDU}}= h(g( \underline{SoC}_{i\text{,}t }^{\text{DIU}}\text{,}c^{\text{CM}})\text{,}D_{i\text{,}t})\text{, }\forall t\in \mathcal{T}_{\text{E}}^{(k)}\text{, }\forall i\in \mathcal{B} \label{DIU-DDUup}\\
  &\hspace{-0.2cm} g= { \underline{SoC}_{i\text{,}t}^{\text{DIU}}}(1-\mathcal{G}({\mu}_{g}\text{,}{\sigma}_{g}))\text{, }\forall t\in \mathcal{T}_{\text{E}}^{(k)}\text{, }\forall i\in \mathcal{B} \label{spg}\\
 &\hspace{-0.2cm} h= ( \underline{SoC}_{i\text{,}t}^{\text{B}}-Q_{g})\mathcal{H}({\mu}_{h}\text{,}{\sigma}_{h})+Q_{g}\text{, }\forall t\in \mathcal{T}_{\text{E}}^{(k)}\text{, }\forall i\in \mathcal{B}\label{sph}\\
 &\hspace{-0.2cm} {\mu}_{g}=\alpha_{i}c_{t}^{\text{CM}}\text{, }{\mu}_{h}={\beta }_{i}D_{i\text{,}t}\text{, }\forall t\in \mathcal{T}_{\text{E}}^{(k)}\text{, }\forall i\in \mathcal{B}\label{RDIN}\\
  &\hspace{-0.2cm} D_{i\text{,}t} =\dfrac{\rho \sum\nolimits_{\kappa=1}^{k-1}{D}_{i\text{,}t_{\text{e}}^{(\kappa)}}}{k-1}+(1-\rho )\times\{\lambda \sum\nolimits_{\tau =t_{\text{s}}^{(k)}}^{t} \dfrac{P_{\text{d}\text{,}i\text{,}\tau }^{\text{RD}}}{{\overline{P}_{\text{d}\text{,}i\text{,}\tau }}T}\label{RDfunction} \\
  &\hspace{0.4cm} +(1-\lambda )|SoC_{i\text{,}t}^{\text{RD}}- SoC_{i\text{,}t}^{\text{B}}|\}\text{, }\forall t\in \mathcal{T}_{\text{E}}^{(k)}\text{, }\forall i\in \mathcal{B} \nonumber
\end{align}
\end{subequations}
\noindent where the set for time slots of each capacity market call denotes $\mathcal{T}_{\text{E}}^{(k)}=\{t_{\text{s}}^{(k)}\text{,}t_{\text{s}}^{(k)}+1\text{,}\cdots\text{,}t_{\text{e}}^{(k)}\}\text{,} k=1\text{,}2\text{,}\cdots\text{,}M$. The superscript $k$ denotes the number of capacity market calls. Continuous decision variables for charge power, discharge power, SoC, and discomfort in re-dispatch denote $P_{\text{c}\text{,}i\text{,}t}^{\text{RD}}$, $P_{\text{d}\text{,}i\text{,}t}^{\text{RD}}$, $SoC_{i\text{,}t}^{\text{RD}}$, and $D_{i\text{,}t}$. Discrete decision variables for charge and discharge states in re-dispatch denote $S_{\text{c/d}\text{,}i\text{,}t}^{\text{RD}}$. Parameters are defined as follows, which can either be obtained from
system settings or learned distribution. $\overline{SoC}_{i\text{,}t}^{\text{DIU}}$ and $\underline{SoC}_{i\text{,}t}^{\text{DIU}}$ denote the upper and lower SoC bounds under DIU.  $\overline{SoC}_{i\text{,}t}^{\text{DDU}}$ and $\underline{SoC}_{i\text{,}t}^{\text{DDU}}$ denote the upper and lower SoC bounds under DDU. $\epsilon$ is the probability level of the chance-constraint. The quantile value of distribution $g$ denotes $Q_{g}$. The weight coefficient for discomfort function denote~$\rho\text{,}\ \lambda$ ($0 \le\rho\text{,}\ \lambda \le 1$). The average baseline performance of SoC denotes $SoC_{i,t}^{\text{B}}$ and is equal to the middle value of the SoC bounds under DIU.
\begin{remark}[Learning of DDU Structure]
In practice, the system operator can leverage historical response performance and collaborate with utility companies to learn the parameters and distributions of DDU. Although learning DDU has become a timely topic~\cite{cheng2021operational,kane2020data,ellinas2024decision}, accurately understanding DDU remains a challenge due to the lack of data. Therefore, we provide a tractable distributional robust chance-constrained approach in section~\ref{solution} to handle DDU based on the ambiguous information of DDU. And we further discuss the impact of DDU structure in Section~\ref{DDUstructure}.
\end{remark}
\begin{remark}[Re-dispatch Strategy] Previous works normally employ adequacy-oriented dispatch approach (e.g., fixed dispatch and greedy dispatch). The adequacy-oriented dispatch only addresses GES operations during emergency states, while overlooking cross-market strategic behavior. In contrast, market-oriented dispatch is more realistic and can coordinate the price-arbitrage in the energy market with the capacity provision in the capacity market. However, existing works focus on uninterruptible power supply storage~\cite{yong2023capacity} and centralized storage~\cite{dratsas2023real}, where storage can be treated as deterministic. As illustrated in Fig.~\ref{DDUbounds}, applying these methods to GES ensures feasibility under theoretical DIU bounds but fails to meet the practical DDU bounds, which may result in significant response unavailability in the contraction stage (from $t_{\text{m}}$ to $t_{\text{e}}$). Therefore, we propose a chance-constrained optimization to explicitly address the inherent DDU in GES. And we also propose a practical GES performance simulation in Section~\ref{consequence} to reveal the consequence of overlooking DDU.
\end{remark}

\textbf{(iii) Recovery State:} GES will adopt a rule-based decision policy as~\eqref{recoverstate} to transition back to the pre-dispatch strategy. The charge or discharge power of GES should be limited with power bounds, SoC bounds and system net capacity.
\begin{subequations}\label{recoverstate}
\begin{align}
 P_{\text{c}\text{,}i\text{,}t}^{\text{RD}}=&\min \{[SoC_{i\text{,}t}^{\text{PD}}-(1-{{\varepsilon }_{i}}\Delta t)SoC_{i\text{,}t-1}^{\text{RD}}]{{E}_{i}}/({{\eta}_{\text{c},i}}\Delta t) \\\nonumber
& {{\overline{P}}_{\text{c}\text{,}i\text{,}t}}\text{, } {{\varphi }_{i\text{,}t}}NC_{t}\}\text{, }\forall t\in \mathcal{T}_{\text{R}}^{(k)}\text{, }\forall i\in \mathcal{B} \\
P_{\text{d}\text{,}i\text{,}t}^{\text{RD}}=&\min \{[(1-{{\varepsilon }_{i}}\Delta t)SoC_{i\text{,}t-1}^{\text{RD}}-SoC_{i\text{,}t}^{\text{PD}}]{{E}_{i}}{{\eta}_{\text{d}\text{,}i}}/\Delta t\\\nonumber
& {{\overline{P}}_{\text{d}\text{,}i\text{,}t}}\}\text{, }\forall t\in \mathcal{T}_{\text{R}}^{(k)}\text{, }\forall i\in \mathcal{B}  
\end{align}
\end{subequations}
where the set for time slots of recovery states denotes $\mathcal{T}_{\text{R}}^{(k)}=\{t_{\text{e}}^{(k)}+1\text{,}t_{\text{e}}^{(k)}+2\text{,}\cdots\text{,}t_{\text{r}}^{(k)}\}\text{,} k=1\text{,}2\text{,}\cdots\text{,}M$. ${{\varphi }_{i\text{,}t}}={{\overline{P}}_{\text{c}\text{,}i\text{,}t}}/\sum\nolimits_{i}{{\overline{P}}_{\text{c}\text{,}i\text{,}t}}$ denotes the capacity allocation ratio of GES at bus \textit{i}.

\subsection{Problem Reformulations and Solution Methodology}~\label{solution}
Chance constraints~\eqref{DDU}-\eqref{RDfunction}~admit a compact form~\eqref{cf}. ${{a}_{i}}{(\bm{x})}\text{,}{{b}_{i}}(\bm{x})$ are affine functions of decision set $\bm{x}$. $\bm{\xi}(\bm{x})$ denotes the set of DDU with mean $\bm{\mu}(\bm{x})$ and covariance $\bm{\Sigma}(\bm{x})$. We have the deterministic reformulation of~\eqref{cf}~as~\eqref{drf}. While DDU couples the inverse cumulative distribution function~$F_{\bm{x}}^{-1}$~with decisions~$\bm{x}$, rendering it unknown prior to optimization and thus making the problem intractable to solve. In previous works~\cite{qi2023chance,qi2023portfolio,pan2022modeling}, the DDU is modeled either using ambiguous or complete knowledge of the uncertainty distributions. However, data scarcity makes it difficult to accurately characterize the complete distribution of DDU, and using the robust method~\cite{su2022multi} may lead to overly conservative results. Hence, we further propose a data-driven reformulation in~\eqref{ddr} based on the acquired data and information of DDU. We use the available data and distributionally robust value in Table~\ref{approximation} to approximately substitute the DDU term in the reformulation~\eqref{drf}. The approximation accuracy improves with the amount of data and DDU information available. Specifically, if no data regarding DDU is available, one can select the normalized inverse cumulative distribution value corresponding to the first entry (NA) in the table. Alternatively, if the obtained data indicates a unimodal distribution, the value corresponding to the third entry (U) in the table should be selected. When the data quantity \( K \rightarrow \infty \), the reformulation in~\eqref{ddr} asymptotically approaches~\eqref{cf}. We defer the complete proof to Appendix~\ref{methodproof}.
\begin{subequations}
\begin{align}
   & \mathbb{P}\left( {{a}_{i}}{(\bm{x})}^{\text{T}}\bm{\xi}(\bm{x}) \le {{b}_{i}}(\bm{x}) \right) \ge 1-\epsilon \label{cf}\\
   & {{a}_{i}}{(\bm{x})}^{\text{T}}\bm{\mu}(\bm{x})+F^{-1}_{\bm{x}}(1-\epsilon)\sqrt{{{a}_{i}}{(\bm{x})}^{\text{T}}\bm{\Sigma}(\bm{x}){{a}_{i}}{(\bm{x})}} \le {{b}_{i}}(\bm{x})  \label{drf}
\end{align}   \vspace{-0.5cm}
\end{subequations}
\begin{subequations}\label{ddr}
\begin{align}
   &  {a}_{i}{(\bm{x})}^{\text{T}}\bm{\mu}(\bm{x})+\psi_K\bm{r}(\bm{x})+\pi_{K}\overline{F}^{-1}_{\bm{x}}(1-\epsilon)\parallel\bm{y}\parallel_2 \le {{b}_{i}}(\bm{x}) \\
    & \sqrt{{{a}_{i}}{(\bm{x})}^{\text{T}}\bm{\Sigma}{{a}_{i}}{(\bm{x})}} \le y_{1}\text{,}\sqrt{2\psi_K}\bm{r}(\bm{x}) \le y_{2} \\
    & \psi_K=K^{(1/p-1/2)} \\
    &\pi_{K}={{( 1-\dfrac{4}{\epsilon }\exp( -{{( {{K}^{1/p}}-2)}^{2}}/2 ) )}^{-1/2}}
\end{align}
\end{subequations}
where $\bm{r}_{\bm{x}}$~is the radius of DDU set, $\bm{y}$~is the auxiliary decisions. $\overline{F}^{-1}_{\bm{x}}$ is the robust approximation value of the inversed cumulative distribution function, which can be obtained by using generations of Cantelli’s inequality, as shown in Table~\ref{approximation}. $p$ is a constant and $K$ is the number of observed samples, which should guarantee that $p\ge2\text{, }K>{{(2+\sqrt{2\ln (4/\epsilon )})}^{p}}$.

\begin{table}[!ht]
  \centering
  \caption{\rmfamily Robust Approximation of Widely Used Normalized Inverse Cumulative Distribution}
  \setlength{\tabcolsep}{0mm}{
      \begin{tabular}{l c c}
    \toprule
    Type \& Shape & $F^{-1}_{\bm{x}}(1-\epsilon)$ & $\epsilon$ \\
    \midrule
    1) No distribution assumption (NA) & $\sqrt{(1-\epsilon )/\epsilon }$ & $0<\epsilon \le 1$ \\\specialrule{0em}{0.5em}{0em}
    \multirow{2}[0]{*}{2) Symmetric distribution (S)} & $\sqrt{1/2\epsilon }$ & $0<\epsilon \le 1/2$ \\\specialrule{0em}{0.5em}{0em}
          & 0     & $1/2<\epsilon \le 1$ \\\specialrule{0em}{0.5em}{0em}
    \multirow{2}[0]{*}{3) Unimodal distribution (U)} & $\sqrt{(4-9\epsilon )/9\epsilon }$ & $0<\epsilon \le 1/6$ \\\specialrule{0em}{0.5em}{0em}
          & $\sqrt{(3-3\epsilon )/(1+3\epsilon )}$ & $1/6<\epsilon \le 1$ \\\specialrule{0em}{0.5em}{0em}
    \multirow{3}[0]{*}{\parbox{3cm}{4) Symmetric \& unimodal distribution (SU)}} & $\sqrt{2/9\epsilon }$ & $0<\epsilon \le 1/6$ \\\specialrule{0em}{0.5em}{0em}
          & $\sqrt{3}(1-2\epsilon )$ & $1/6<\epsilon \le 1/2$ \\\specialrule{0em}{0.5em}{0em}
          & 0     & $1/2<\epsilon \le 1$ \\\specialrule{0em}{0.5em}{0em}
    \bottomrule
    \end{tabular}%
    }
  \label{approximation}%
\end{table}%

After the reformulation, the problem~\eqref{emergencystate} is trivial to solve with existing commercial solvers. The proposed market-oriented risk-averse redispatch is summarized in Algorithm~\ref{algorithm1}.

\begin{algorithm}[htbp]\label{algorithm1}
\hspace*{\fill}
\caption{Market-Oriented Risk-Averse Coordinated Dispatch}
\SetAlgoLined
\SetAlgoNoEnd 
\KwIn{Setup, net capacity $NC_{t}$, probability level $\epsilon$.}
\KwOut{GES re-dispatch strategy~$\{P_{\text{c/d}\text{,}i\text{,}t}^{\text{RD}}\text{,}\  SoC_{i\text{,}t}^{\text{RD}}\}$.}
\SetKwBlock{StepOne}{Step 1 - Pre-dispatch}{}
\SetKwBlock{StepTwo}{Step 2 - Re-dispatch}{}

\StepOne{\textbf{do} optimization~\eqref{normalstate} to generate pre-dispatch\\ strategy of each GES, i.e., $\{P_{\text{c/d}\text{,}i\text{,}t}^{\text{PD}}\text{,}\  SoC_{i\text{,}t}^{\text{PD}}\}$.}

\StepTwo{
    \For{$t=1$ \KwTo $NT$}{
        \If{${{NC}_{t}} \geq 0$}{
            \If{$\Delta SoC_{i\text{,}t} = 0$}{
                $\{P_{\text{c/d}\text{,}i\text{,}t}^{\text{RD}}\text{,}\  SoC_{i\text{,}t}^{\text{RD}}\} \leftarrow \{P_{\text{c/d}\text{,}i\text{,}t}^{\text{PD}}\text{,}\  SoC_{i\text{,}t}^{\text{PD}}\}$. 
            }
            \Else{
                \textbf{do} rule-based decision policy~\eqref{recoverstate} to \\ generate $\{P_{\text{c/d}\text{,}i\text{,}t}^{\text{RD}}\text{,}\  SoC_{i\text{,}t}^{\text{RD}}\}$.
            }
            \textbf{end}
        }
        \Else{
            \textbf{do} optimization~\eqref{emergencystate} with reformulation in\\ Table~\ref{approximation} to generate $\{P_{\text{c/d}\text{,}i\text{,}t}^{\text{RD}}\text{,}\  SoC_{i\text{,}t}^{\text{RD}}\}$.
        }
        \textbf{end}
    }
}
\end{algorithm}
\subsection {Practical GES Performance Simulation}~\label{consequence}
To evaluate the practical response of GES to the dispatch strategy and reveal the consequences of overlooking DDU, we present a practical GES performance simulation following the steps outlined in Algorithm~\ref{algorithm2}. The difference between the practical SoC bounds and the theoretical ones is the primary reason for the (increased) response unavailability of GES. Therefore, we first calculate the practical SoC bounds by substituting theoretical strategy ($P_{\text{c/d}\text{,}i\text{,}t}^{\text{T}}$) generated either by the proposed method or by the existing dispatch methods, into DDU model~\eqref{DIU-DDUup}-\eqref{RDfunction}. If the theoretical SoC strategy is outside the practical SoC bounds, it represents a 100\% reliable response. However, if it falls within the practical SoC bounds, it indicates response unavailability. Based on this principle, we update the practical GES operational strategy ($P_{\text{c/d}\text{,}i\text{,}t}^{\text{P}}$).

As shown in Fig.~\ref{DDUbounds}, from the start of capacity market call (from $t_{\text{s}}$ to $t_{\text{m}}$),  it is observed that, when overlooking DDU, the practical SoC bounds (green rainbow) expand compared with theoretical ones as the incentive effect dominates. However, from $t_{\text{m}}$ to $t_{\text{e}}$, the practical SoC bounds (red rainbow) contract compared to the theoretical ones as the discomfort effect takes over, leading to response unavailability during this period. In contrast, the proposed method effectively ensures that the SoC strategy can meet the practical SoC bounds, thus guaranteeing the reliability of the GES response.
\begin{algorithm}[htp]\label{algorithm2}
\hspace*{\fill}
\caption{Practical GES Performance Simulation}
\SetAlgoLined
\SetEndCharOfAlgoLine{}
\KwIn{Theoretical GES strategy~$\{P_{\text{c/d}\text{,}i\text{,}t}^{\text{T}}\text{,} SoC_{i\text{,}t}^{\text{T}}\}$.}
\KwOut{Practical GES strategy~$\{P_{\text{c/d}\text{,}i\text{,}t}^{\text{P}}\text{,} SoC_{i\text{,}t}^{\text{P}}\}$.}
\SetKwBlock{StepOne}{Step 1 - Practical SoC Bounds Simulation}{}
\SetKwBlock{StepTwo}{Step 2 - Practical GES Operations Updates}{}
\SetKw{Parallel}{parallel}
\StepOne{\For{$k=1$ \KwTo $M$}{
\For{$t=t_{\text{s}}^{(k)}$ \KwTo $t_{\text{e}}^{(k)}$}{Calculate the practical SoC bound under \\ DDU by DDU model~\eqref{DIU-DDUup}-\eqref{RDfunction}  and\\ theoretical GES strategy $\{P_{\text{c/d}\text{,}i\text{,}t}^{\text{T}}\text{,} SoC_{i\text{,}t}^{\text{T}}\}$.\;
}
}
}
\StepTwo{\For{$k=1$ \KwTo $M$}{
\For{$t=t_{\text{s}}^{(k)}$ \KwTo $t_{\text{e}}^{(k)}$}{(i) Update practical SoC strategy of GES\\  ${SoC}_{i\text{,}t}^{\text{P}}$, by restricting the theoretical SoC\\  strategy ${SoC}_{i\text{,}t}^{\text{T}}$ within the practical lower\\ SoC bound; (ii) Calculate practical\\ strategy of GES by~\eqref{soc-powerR}, i.e., ${P}_{\text{c/d}\text{,}i\text{,}t}^{\text{P}}$~.\;
}
}
}
\end{algorithm}

\section{CC Evaluation Framework of GES}\label{evaluation}

CC evaluation of GES is implemented by evaluating and comparing the system adequacy for the cases with and without GES. A detailed explanation of each evaluation step can be found in~\cite{zeng}. Compared with the traditional CC evaluation procedures under DIU, our additional methodological contributions under DDU are explicitly highlighted in yellow within Fig.~\ref{flowchart}. The key differences between these two frameworks are briefly outlined as follows:
\begin{figure}[!t]
        \setlength{\abovecaptionskip}{-0.1cm}  
    \setlength{\belowcaptionskip}{-0.1cm} 
      \begin{center}
\includegraphics[width=1\columnwidth]{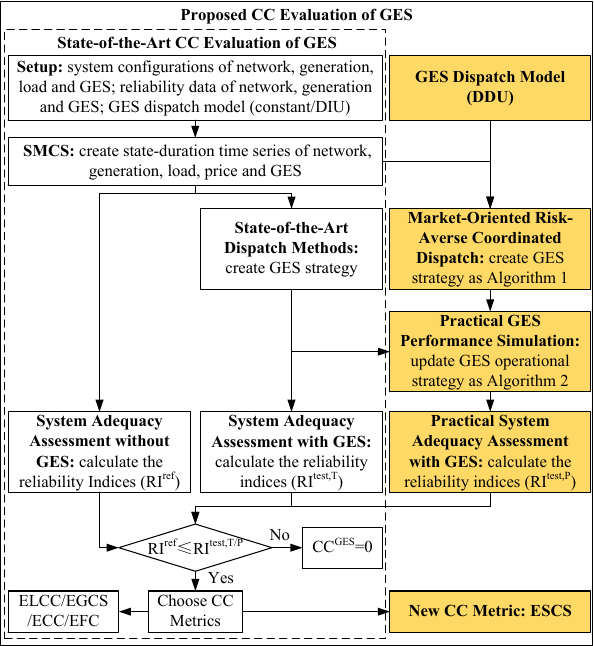}
    \caption{Flowchart of CC evaluation of GES under DDU.}\label{flowchart}
  \end{center}\vspace{-0.5cm}
\end{figure}
\begin{enumerate}
    \item \textbf{Setup:} CC evaluation requires data collection on system configurations for the network, generation, load, and GES; reliability data for network, generation, and GES; and GES dispatch model parameters. State-of-the-art methodology consider DIU in random failures, stochastic renewable output, and stochastic response of GES. In contrast, the proposed methodology further addresses DDU in the available capacity of GES during capacity market operation, which is modeled in Section~\ref{dispatch}.
    \item \textbf{SMCS:} SMCS is a fundamental technique used in adequacy assessment~\cite{zeng}, which creates time-duration series of random failures and uncertainties over a long-term horizon to guarantee the convergence. SMCS can maintain the inter-temporal information of samples which is beneficial to evaluate system adequacy with GES. The operational state of generation, lines, and GES is sampled based on a two-state Markov chain. In addition, the profiles of renewable output, load, price and baseline consumption of GES can be generated using either historical or synthetic data~\cite{li2024long,fu2025knowledge}. 
    \item \textbf{GES Dispatch Simulation:} Previous works typically adopt adequacy-oriented dispatch methods, which misalign with the profit-maximization goals of GES and overlook the DDU of GES. Instead, we propose a market-oriented risk-averse coordinated dispatch method to capture the realistic cross-market operations of GES and mitigate the risks posed by DDU. Additionally, we simulate the practical GES performance and address the GES response unavailability due to DDU under different theoretical dispatch strategies. This part is detailed in Section~\ref{dispatch}.
    \item \textbf{System Adequacy Assessment:} Reliability indices, such as expected energy not served (EENS), are used to quantify system adequacy and are updated within the SMCS until the convergence criterion is satisfied. Previous adequacy assessments only consider loss of load events from DIU, assuming all resources will respond as dispatched. The results are regarded as theoretical and ex-ante, as defined in~\eqref{reliabilityT}. In contrast, the proposed adequacy assessment additionally considers loss of load events from DDU and addresses the practical performance of GES affected by DDU, as shown in~\eqref{reliabilityP}. And the convergence criterion, i.e., coefficient of variance (CoV) is defined in~\eqref{criterion}. Loss of load events from DIU and DDU are defined as ${X}_{t}$ and ${Y}_{t}$, respectively. $\bm{x}$ and $\bm{\xi}$ are sets of decisions and uncertainties, respectively. $P_{t}^{\text{LC}}$ represents load curtailment power. $\sigma$ and $E$ are functions of standard deviation and expectation.
\begin{subequations}
    \begin{align}
&\text{EENS}_{t}^{\text{T}}\text{=}(8760/t)\cdot\sum\nolimits_{\tau=1}^{t}{P_{t}^{\text{LC}}({{X}_{t}}|\bm{\xi})}\label{reliabilityT}\\
&\text{EENS}_{t}^{\text{P}}\text{=}(8760/t)\cdot\sum\nolimits_{\tau=1}^{t}{P_{t}^{\text{LC}}({{X}_{t}}\text{,}{{Y}_{t}}|\bm{x}\text{,}\bm{\xi})}\label{reliabilityP}\\
& \text{CoV}=\sigma(\text{EENS}_{t})/{ E(\text{EENS}_{t})} \leq 5\% \label{criterion}
\end{align}
\end{subequations}
    \item \textbf{CC Metrics Calculation:} Each existing CC metric is defined for different planning scenarios and goals, representing how much generation or load capacity can be substituted in system planning or capacity market pricing. For instance, EFC and ECC quantify the substitution of generation capacity expansion, with EFC assuming perfect generation without failures. EGCS and ELCC quantify the substitution of retired generation capacity and load-carrying capacity, respectively. However, most GES resources operate more like ES by charging and discharging to contribute to resource adequacy. Thus, to maintain the inter-temporal information and address long-duration power shortages, it is more effective to replace the stochastic GES with an equivalent deterministic ES. This approach significantly enhances computational efficiency when integrating stochastic GES into the co-planning of transmission systems and storage~\cite{gan2019security}. Hence, this paper introduces a CC metric to quantify the equivalent deterministic storage capacity of the uncertain GES. The calculation of ESCS is given by~\eqref{ESCS}. $\text{R}{{\text{I}}^{\text{ref}}}$ and $\text{R}{{\text{I}}^{\text{test}}}$ denote the reliability of the referenced and tested system, respectively. ${{C}^{\text{CG}}}$, ${{C}^{\text{RG}}}$, and ${{C}^{\text{LD}}}$ are rated capacity of conventional generation, renewable generation and load. ${{C}^{\text{RP}}}$~and~${{C}^{\text{GES}}}$~are the replaced capacity of deterministic ES and rated power capacity of GES. $R$ is the reliability function.
\begin{subequations}
\begin{align}
& \text{R}{{\text{I}}^{\text{ref}}}=R[{{C}^{\text{CG/RG}}}\text{; }{{C}^{\text{ES}}}+{{C}^{\text{RP}}}\text{; }C^{\text{LD}}]\\
&\text{R}{{\text{I}}^{\text{test}}}=R[{{C}^{\text{CG/RG}}}\text{; }{{C}^{\text{ES}}}+{{C}^{\text{GES}}}\text{; }C^{\text{LD}}] \\ 
 & \text{R}{{\text{I}}^{\text{ref}}}=\text{R}{{\text{I}}^{\text{test}}}\text{, }\text{ESCS}=\text{{C}}^{\text{RP}}/\text{{C}}^{\text{GES}}
\end{align}\label{ESCS}
\end{subequations}    
\end{enumerate}


\section{Numerical Studies}~\label{case}
\subsection {Set-Up}
The proposed method is tested in the modified IEEE RTS-79 benchmark system~\cite{zhou2016}. The test system includes 32 conventional generators with a total installed capacity of 3450 MW and load capacity of 2850 MW with winter peak.
Compared with the original system, we further introduce the RG, ES, and VES into the system. The RG, consisting of equal-capacity solar and wind, are located at PV buses and contribute a share of the total generation. Front-of-meter (FTM) ES resources are located on PV buses, while behind-of-meter (BTM) ES resources are located on PQ buses. VES resources are located at PQ buses. The grid reliability data are inherited from the original system. 20 years of historical data on renewable generation (RG) power, load, and energy market prices were collected from the Belgian TSO~\cite{Elia-data} for SMCS. The capacity market price sets at \$2000/MW per month~\cite{PFP}. The optimization is coded in MATLAB with the YALMIP interface and solved by
GUROBI 11.0 solver. The programming environment is Core i9-13900HX @ 2.20GHz with RAM 32 GB.

\subsection {GES Operation and Adequacy Performance}

We first demonstrate the performance of the proposed method in terms of GES operation and adequacy performance, compared with state-of-the-art methods.
\begin{enumerate}
    \item \textit{(M1) Fixed Dispatch}: GES operation under fixed dispatch~\cite{zhou2016,safdarian} is implemented during peak days (i.e., daily peak load over 90\% of the annual peak load). The optimization aims to minimize the annual peak using a deterministic approach determined prior to SMCS. 
    \item \textit{(M2) Greedy Dispatch}: Under greedy dispatch~\cite{evans2019}, GES charges during normal system state and discharges during capacity deficiencies. GES operation is determined within SMCS using a rule-based policy.
    \item \textit{(M3) Re-dispatch}: Under re-dispatch~\cite{yong2023capacity}, GES follows a pre-dispatch strategy for price arbitrage during normal system state. While during emergencies, GES is re-dispatched to discharge, and then reverts to the pre-dispatch strategy after emergencies. The pre-dispatch optimization is determined prior to SMCS, whereas the re-dispatch strategy is formulated within SMCS.
    \item \textit{(M4) Risk-Averse coordinated Dispatch}: Compared to~\textbf{M3}, the proposed approach incorporates an minimum capacity withholding ($\underline{SoC}^{\text{CW}}$=0.2) for pre-dispatch optimization. Moreover, we introduce distributional robust chance constraints ($\epsilon$=5\%) to address DDU in the re-dispatch optimization.
\end{enumerate}

\begin{figure}[!ht]
    \setlength{\abovecaptionskip}{-0.1cm}  
    \setlength{\belowcaptionskip}{-0.1cm} 
  \begin{center}
\includegraphics[width=1\columnwidth]{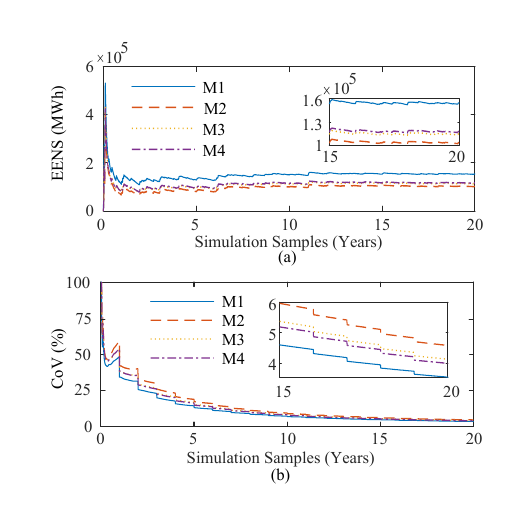}
    \caption{Convergence performance of SMCS under different dispatch methods: (a) EENS and (b) CoV. }\label{convergenceperformance}
  \end{center}\vspace{-0.5cm}
\end{figure}

First, 4-hour FTM-ES is tested with the following settings: forced outage rate of 5\%, MTTR of 24 hours, self-discharge rate of 5\% per month, one-way efficiency of 90\%, and power rating of 30\% of the RG rated power. The RG penetration is set to 30\%. Fig.~\ref{convergenceperformance} illustrates the convergence process of both EENS and CoV within the SMCS. It can be observed that after 20 years of simulation (i.e., 175200 samples), the EENS metric stabilizes, and the convergence criterion drops below 5\% for all dispatch methods. Therefore, in the subsequent analysis, we utilize the simulated results from the 20-year SMCS. We employ parallel computing techniques to accelerate the non-sequential processes within SMCS, with the average computation times for M1-M4 being 59.6s, 0.4s, 273.0s, and 1354.0s, respectively. The longer computation time for M4 is primarily due to the additional calculations required for pre-dispatch and the optimization under DDU. Although CC evaluation is typically conducted offline with relatively relaxed computational requirements, practical applications can still benefit from scalable computational approaches. For instance, aggregated GES models, parallel computing for pre-dispatch strategies, and simplified DC-OPF formulations can all be employed to significantly enhance scalability.

\begin{figure}[!ht]
      \setlength{\abovecaptionskip}{-0.1cm}  
    \setlength{\belowcaptionskip}{-0.1cm} 
  \begin{center}
\includegraphics[width=1\columnwidth]{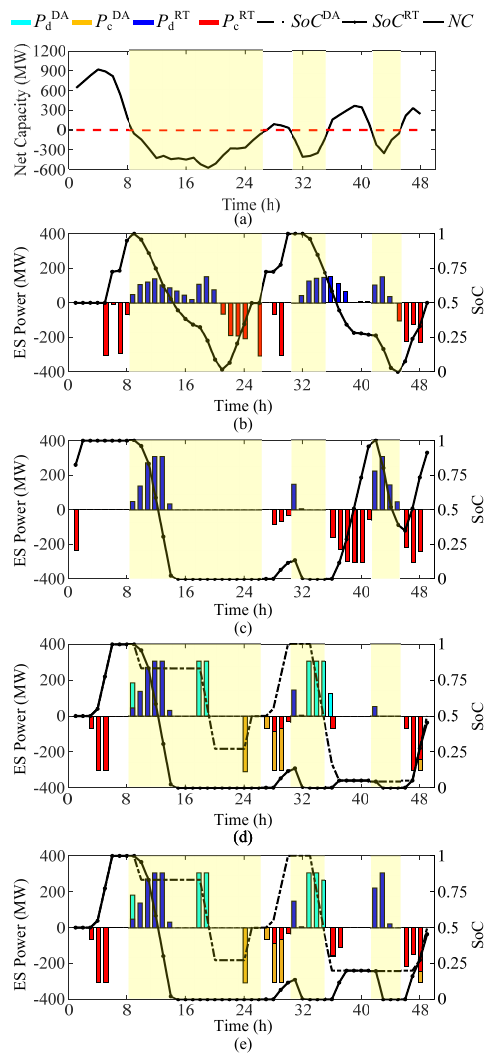}
    \caption{Simulations of two consecutive days: (a) net capacity and ES operations under different dispatch methods: (b) M1, (c) M2, (d) M3, and (e) M4.}\label{gesoperaions}
  \end{center}\vspace{-0.5cm}
\end{figure}

\begin{figure}[!ht]
      \setlength{\abovecaptionskip}{-0.1cm}  
    \setlength{\belowcaptionskip}{-0.1cm} 
  \begin{center}
\includegraphics[width=1\columnwidth]{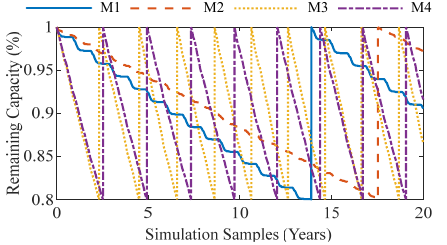}
    \caption{Degradation of ES under different dispatch methods.}\label{degradation}
  \end{center}\vspace{-0.5cm}
\end{figure}

Fig.~\ref{gesoperaions} illustrates the significant differences in storage operations under four dispatch methods. And the average adequacy performance is summarized in Table~\ref{dispatchresult}. It is observed that M1 only addresses capacity deficiencies during peak periods (9h-20h, 32h-44h) but fails to respond to real-time failures at 21h-26h and 31h, even though storage still has remaining SoC, which results in the minimal flexibility utilization and the lowest adequacy performance. While greedy dispatch maintains the full capacity at the normal system state and ES is only discharged in response to all the loss of load events, ending with ``0'' SoC. Although it effectively addresses real-time failures and outperforms other methods in terms of adequacy performance, it is only suitable for centralized (non-merchant) storage, which has no capacity to provide other services beyond resource adequacy. Compared to two adequacy-oriented dispatch methods (M1\&M2), market-oriented dispatch (M3\&M4) can capture more realistic GES operations across multiple markets. They involve both a pre-dispatch strategy for price arbitrage (light-blue and light-orange bars) and a re-dispatch strategy for minimizing loss of load (blue and red bars). While, the proposed method (M4) further simulates ES capacity withholding to maintain the SoC above 20\% and addresses DDU during emergency states to limit ES response unavailability. This results in increased discharging during real-time failures and improved practical adequacy performance compared to M3. Additionally, we illustrate the degradation of ES in Fig.~\ref{degradation}. In line with existing industrial practice, the ES is retired once its remaining capacity falls below 80\%. It can be observed that, compared to methods M1 and M2, the remaining capacity of ES under methods M3 and M4 declines more rapidly due to frequent switching and provision of services between the energy and capacity markets. Moreover, compared with M3, the degradation rate in M4 is slower because DDU is explicitly considered, resulting in fewer replacements (one less battery replacement). 

The difference in storage operations is the primary cause of the variations in adequacy performance. As is shown in Table~\ref{dispatchresult}, the theoretical adequacy performance follows the order: M2 $>$ M3 $>$ M4 $>$ M1. However, better theoretical adequacy performance does not guarantee more realistic storage operations. Regarding uncertainty considerations and their consequences, we found that the practical adequacy performance of state-of-the-art methods is worse than the theoretical performance, as they overlook DIUs (M1-M2) or DDUs (M1-M3) in GES. After simulating practical GES performance, for 30\% rated power 4-hour FTM ES, practical EENS increased by 1.2\%, 6.1\%, and 8.0\% using the M1-M3 methods, respectively. The additional practical EENS increases as either the power or energy capacity rises, and for 50\% rated power 12-hour ES, the increase of EENS under M3 reaches 20.5\%. By contrast, the proposed method effectively controls the risk from DDU within the probability threshold (5\%), resulting in minimal EENS increase (less than 0.2\%). Consequently, method M4 achieves the best practical adequacy performance, whereas methods M1 and M3 perform approximately 40\% and 10\% worse, respectively. The consequence of overlooking DDU is not significant for ES, as it primarily stems from long-term capacity degradation. However, the situation will be much worse for VES.

\begin{table}[!ht]
      \setlength{\abovecaptionskip}{-0.1cm}  
    \setlength{\belowcaptionskip}{-0.1cm} 
  \centering
  \caption{\rmfamily Adequacy Performance Compared with Different Dispatch Methods and Storage Configurations}
    \setlength{\tabcolsep}{0.5mm}{
    \begin{tabular}{cccccc}
    \toprule
    \multirow{2}[4]{*}{\makecell{Dispatch\\ Methods}} & \multirow{2}[4]{*}{\makecell{Rated\\ Power}} & 4-h-ES & 4-h-VES & 12-h-ES & 12-h-VES \\
\cmidrule{3-6}          &       & \makecell{$\text{EENS}^{\text{T/P}}$\\(GWh)} & \makecell{$\text{EENS}^{\text{T/P}}$\\(GWh)} & \makecell{$\text{EENS}^{\text{T/P}}$\\(GWh)} & \makecell{$\text{EENS}^{\text{T/P}}$\\(GWh)} \\
    \midrule
    \multirow{2}[1]{*}{M1} & 30\%  & 154.7/156.6  & 162.2/164.0  & 139.7/141.0  & 153.2/160.6  \\
          & 50\%  & 151.1/152.7  & 155.8/159.6  & 137.6/137.7  & 149.2/153.2  \\
    \multirow{2}[0]{*}{M2} & 30\%  & 103.4/108.7  & 129.9/151.1  & 65.3/71.1  & 120.6/142.9  \\
          & 50\%  & 84.2/90.2  & 121.4/143.5  & 45.9/52.1  & 116.6/135.5  \\
    \multirow{2}[0]{*}{M3} & 30\%  & 114.6/120.9  & 136.4/145.4  & 73.5/82.7  & 121.3/136.4  \\
          & 50\%  & 96.2/106.4  & 126.4/135.8  & 55.9/67.3  & 117.0/126.2  \\
    \multirow{2}[1]{*}{M4} & 30\%  & 112.5/112.8  & 141.2/141.3  & 73.4/73.5  & 128.6/128.7  \\
          & 50\%  & 93.8/94.0  & 131.9/131.9  & 55.7/55.8  & 122.1/122.2  \\
    \bottomrule
    \end{tabular}%
    }\label{dispatchresult}
\end{table}%
We further compare the theoretical and practical 4-hour VES operations using M3-M4 in Fig.~\ref{DDUdispatch}. We observe that the DDU associated with TCL-VES follows a unimodal distribution, with the corresponding DDU parameters set as $\alpha=1$, $\beta=4$, $\rho=0.2$, $\lambda=0.5$.  In the subsequent case studies, we adopt this form of DDU unless explicitly stated otherwise. It is observed that VES discharges at high power for prolonged periods in M3 to address capacity deficiencies. This leads to significant response discomfort for VES occupants, resulting in extensive response unavailability, as shown by the gap between the theoretical discharge (light-blue bars) and the practical discharge (blue bars) from 12h to 22h and 36h to 46h. By contrast, the proposed method (M4) adopts a smarter strategy, where high-power discharge and low-power discharge or charging alternate in sequence, ensuring the SoC of VES remains within practical bounds. As a result, response unavailability is nearly eliminated. The difference in real-time VES operations primarily lies in the variation in SoC bounds. Compared to DIU bounds in M3 with fixed distribution, the distribution of DDU bounds in M4 is dynamically adjusted based on decisions and explicitly incorporated into the dispatch optimization. And compared to DIU bounds, DDU bounds contract by approximately 45.3\% during emergency states for discharging and expand by about 60.5\% during recovery states for charging. 

Regarding adequacy performance, Table~\ref{dispatchresult} shows that ES outperforms VES, as VES experiences greater energy losses, lower on-state probability, and higher sensitivity to DDU.  We also find that EENS decreases with the increased power or energy capacity. When accounting for the consequence of overlooking DDU, the highest increase in practical EENS, approximately 17.3\%, is observed for 12-hour VES under greedy dispatch. This is due to the extended periods of discharging, which accumulates discomfort and worsens response availability. Furthermore, although M3 employs market-oriented re-dispatch, the overlooking of DDU still results in an 8.2\% increase in practical EENS. While, the proposed method effectively mitigates the risk from DDU, resulting in less than a 0.1\% increase in practical EENS. The proposed method still achieves the best practical adequacy performance, whereas methods M1-M3 perform approximately 15\%, 7\%and 3\% worse, respectively. The above analysis demonstrates the accuracy of the proposed method and how existing methods tend to overestimate the adequacy contribution of GES by overlooking DDU.
\begin{figure}[!ht]
      \setlength{\abovecaptionskip}{-0.1cm}  
    \setlength{\belowcaptionskip}{-0.1cm} 
  \begin{center}
\includegraphics[width=1\columnwidth]{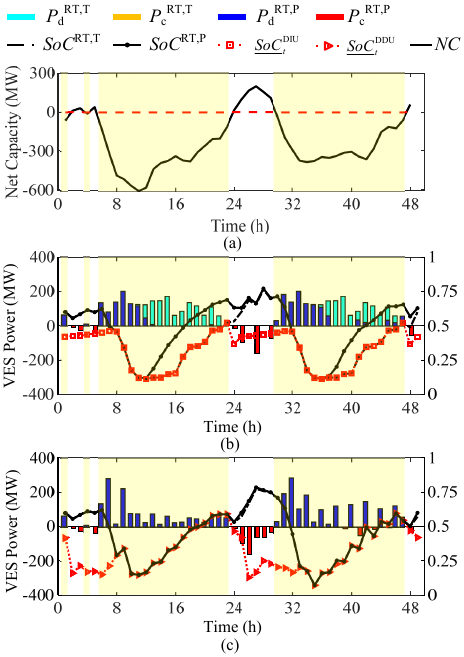}
    \caption{Simulations of two consecutive days: (a) net capacity and VES operations under different dispatch methods: (b) M3 and (c) M4.}\label{DDUdispatch}
 \end{center}\vspace{-0.5cm}
\end{figure}

\subsection {Economic Performance}

Adequacy performance is important for the system operator, but GES participants are more concerned with profit. The profit for GES is primarily composed of three parts: (a) revenue from price arbitrage in the energy market, (b) revenue from capacity provision in the capacity market, and (c) penalty from non-performance in the capacity market. Part (a) is calculated based on~\eqref{objectDA}, part (b) is determined by the reduction in EENS and the value of lost load (VoLL), and part (c) is calculated based on practical performance and penalties for non-performance. According to a common practice in the US~\cite{PFP}, we set VoLL at \$10000/MWh and the penalty for non-performance at \$2000/MWh. We summarize the annual economic performance of 30\% rated power, 4-hour FTM-ES \& 4-hour VES under different dispatch methods in Table~\ref{economic}. With the exception of fixed dispatch (M1), GES can earn substantial revenue in the capacity market, amounting to more than 20 times the revenue from the energy market. Although GES under greedy dispatch (M2) guarantees the highest profit by addressing all real-time capacity deficiencies, it is owned and controlled by the system operator, whose goal is social welfare maximization rather than profit maximization.  Furthermore, GES (especially VES) incurs high penalties (27\% of profit for M2 and 7\% of profit for M3) due to significant response unavailability resulting from the oversight of DDU in the dispatch. This could lead to dissatisfaction and reduce their willingness to participate in the capacity market. Compared to M3, M4 sacrifices approximately \$5.6 million (48\%)/\$3.3 million (44\%) of energy market revenue, representing lost arbitrage opportunities. However, this trade-off results in a corresponding increase in capacity market revenue of \$84.5 million (17\%)/\$41.4 million (16\%), respectively, along with a substantial reduction in penalties. Consequently, the overall profit improves by approximately 18\%.
\begin{table}[!ht]
      \setlength{\abovecaptionskip}{-0.1cm}  
    \setlength{\belowcaptionskip}{-0.1cm} 
  \centering
  \caption{\rmfamily Annual Economic Performance Compared with Different Dispatch Methods and GES Types}
    \setlength{\tabcolsep}{0mm}{
    \begin{tabular}{cccccc}
    \toprule
    \makecell{Dispatch\\ Method} & \makecell{GES\\ Type}  & \makecell{EM Revenue\\ (\$$10^6$)} & \makecell{CM Revenue\\ (\$$10^6$)} & \makecell{CM Penalty\\ (\$$10^6$)} & \makecell{Total Profit\\ (\$$10^6$)} \\
    \midrule
    M1 & ES/VES & 15.5/10.2   & 144.2/70.3  & 3.9/3.5  & 155.9/77.0  \\
   M2 & ES/VES & 0.0/0.0   & 623.5/199.1  & 10.6/42.5   & 612.8/156.6 \\
    M3 & ES/VES & 11.6/7.5  & 501.1/256.3  & 12.6/18.1   & 500.1/245.7   \\
   M4 & ES/VES & 6.0/4.2   & 582.6/297.7   & 0.5/0.0   & 588.2/301.8 \\
            \bottomrule
    \end{tabular}%
    }\label{economic}\vspace{-0.5cm}
\end{table}%

\subsection {Capacity Credit Value and System Decision-Making}

In this section, we demonstrate how the dispatch method and DDU consideration impact the CC value and system decision-making. We focus on average CC values rather than marginal CC values in the following results.

\textbf{(i) CC values across metrics.} We observed slight variations in CC values across different metrics. For instance, for a 30\% rated power, 4-hour FTM-ES under the proposed dispatch method: EFC = 33.5\%, ECC = 36.0\%, ELCC = 28.2\%, and EGCS = 36.2\%. However, these metrics show similar trends across various energy storage configurations and dispatch methods. Therefore, we select ECC as the representative metric for the subsequent analysis.

\begin{figure}[!ht]
      \setlength{\abovecaptionskip}{-0.1cm}  
    \setlength{\belowcaptionskip}{-0.1cm} 
  \begin{center}
\includegraphics[width=1\columnwidth]{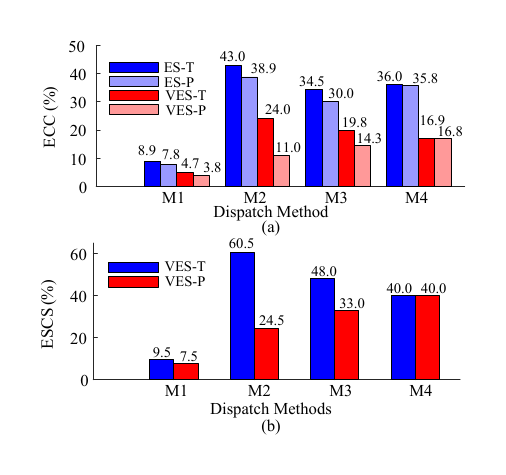}
    \caption{Capacity credit value compared with different dispatch methods: (a) ECC and (b) ESCS.}\label{CCVALUE}
 \end{center}\vspace{-0.5cm}
\end{figure}

\textbf{(ii) CC values under different dispatch method.}
We compare the theoretical (T) and practical (P) CC of 30\% rated power, 4-hour FTM-ES \& VES under different dispatch methods in Fig.~\ref{CCVALUE}. It is observed that the CC of ES ranges from 30\% to 45\% under most dispatch methods except for M1, whereas CC of VES only ranges from 10\% to 25\%. Moreover, under the state-of-the-art methods, overlooking DDU results in a 9.5\%-13.9\% decline in the practical CC of ES, while a more significant decline of 20.8\%-54.2\% is observed for VES. In contrast, with the risk-averse dispatch addressing DDU, the proposed method shows almost no noticeable increase in practical CC. Additionally, without considering DDU, VES is equivalent to at most 60.5\% of ES. However, DDU significantly reduces this capability, resulting in only up to 40\% of ESCS in practice. The above analysis demonstrates that it is essential to adopt a risk-averse dispatch strategy to mitigate DDU risks, thereby enhancing practical CC.

\textbf{(iii) Consequences of inaccurate CC evaluation.} Accurate CC evaluation is crucial for effective system planning and capacity market clearing. When CC is overestimated by 10\%, GES providers initially receive higher committed revenues but may incur financial penalties amounting to approximately 10–15\% of their CM revenues if unable to fulfill capacity obligations in real-time operations. This overestimation suppresses market clearing prices, reducing revenue for other participants and potentially causing underinvestment in reliability-critical resources, ultimately compromising system adequacy. Conversely, underestimating CC by 10\% reduces initial revenue and investment incentives for GES providers by 10\% due to lower contracted capacity. Meanwhile, 5\% higher market clearing prices driven by perceived capacity shortages lead to unnecessary investments from other participants, raising overall system costs. For the system operator, underestimation induces overly conservative planning, resulting in redundant investments and economic inefficiencies on the order of 5–10\%.

\textbf{(iv) System decision-making.} 
Decision-makers should choose the appropriate dispatch method to calculate CC based on their risk preferences and the ownership of the GES. Fixed dispatch methods generate the robust CC of GES, suitable for risk-averse decision-makers. Greedy decision-makers can choose greedy dispatch for centralized (non-merchant) GES. While, market-oriented dispatch strikes a balance between resource adequacy contribution and GES profit, making it more suitable for merchant GES. Additionally, for risk-averse decision-makers, the proposed method should be adopted to mitigate the risks associated with DDU. 

Overall, the advantages of the market-oriented risk-averse re-dispatch method are summarized as follows:
\begin{enumerate}
    \item \textit{Behavior Characterization}: The proposed framework effectively captures the realistic cross-market strategic behavior of GES by incorporating SoC withholding and a market-oriented re-dispatch method. Moreover, the DDU model is introduced to characterize the response willingness of occupants and dynamic availability of GES, which are impacted by incentives and dispatch decisions.
    \item \textit{Practical Adequacy Performance}: By explicitly incorporating DDU in the risk-averse re-dispatch method, the proposed method achieves the highest practical adequacy performance and the lowest response unavailability of GES. This guarantees the highest and accurate CC value when considering DDU. 
    \item \textit{Economic Performance}: The proposed method enhances merchant storage profits by increasing capacity market revenue and reducing penalty, which aligns with the goals of merchant GES. Additionally, using the reliable CC from the proposed method in system planning not only avoids excessive redundancy but also reduces potential costs from power shortages and price spikes.
\end{enumerate}


\section{Sensitivity Analysis}\label{SA}

In this section, we further investigate the key factors impacting the CC of GES using the proposed method.

 \subsection{Impact of GES Configurations on CC}\label{Configurations}
 
  The power and energy capacity significantly affect the CC of GES. Under the 30\% share of RES, the comparative results in Table~\ref{configuration} demonstrate that CC of GES rises with energy capacity, yet falls with increased power capacity. Specifically, the CC of ES with 10\% rated power increases from 42.5\% to 91.3\% as the storage duration extends from 4 hour to 12 hour. While, the CC of 4-hour ES decreases from 42.5\% to 30.4\% as the power capacity increases from 10\% to 50\%. This highlights the critical role of long-duration ES in resource adequacy, and also indicates that CC value diminishes as ES becomes less scarce. Moreover, the CC of VES is not improved significantly by increasing energy capacity (21.8\% for 4-h and 37.1\% for 12-h) due to its inherent discomfort-aversion for long-duration dispatch. 
  
\begin{table}[!ht]
      \setlength{\abovecaptionskip}{-0.1cm}  
    \setlength{\belowcaptionskip}{-0.1cm} 
  \centering
  \caption{\rmfamily Capacity Credit of GES Compared with Different Storage Configurations}
    \setlength{\tabcolsep}{1mm}{
    \begin{tabular}{ccccc}
    \toprule
    \multirow{2}[4]{*}{\makecell{Energy\\ Capacity}} & \multirow{2}[4]{*}{GES Type} & \multicolumn{3}{c}{Power Capacity} \\
\cmidrule{3-5}          &       & 10\%  & 30\%  & 50\%  \\
    \midrule
    4-h & ES/VES    & 42.5\%/21.8\% & 35.8\%/16.8\% & 30.4\%/13.6\%  \\
    8-h & ES/VES     & 73.9\%/31.7\% & 58.2\%/22.4\% & 45.6\%/16.6\%  \\
    12-h & ES/VES   & 91.3\%/37.1\% & 70.3\%/24.8\% & 55.0\%/17.5\%  \\
    \bottomrule
    \end{tabular}
    }\label{configuration}
\end{table}%

 \subsection{Impact of SoC Withholding on CC}\label{SoCWithholding}

The profit and CC of ES are compared with different SoC withholdings in Table~\ref{socwithholding}. For short-duration ES, we observed that the increase in SoC withholding results in a 10\% increase in CC and an 8\% increase in profit. This is due to SoC withholding marginally impacting the energy market revenue while notably boosting the capacity market revenue. For long-duration ES, increasing the SoC withholding leads to a negligible rise of just 1\% in CC and 0.8\% in profit. Hence, SoC withholding should be modeled and considered in the CC evaluation for short-duration ES. And SoC should be monitored and regulated in the capacity market, whereas such regulations are unnecessary for long-duration ES.
\begin{table}[!ht]
      \setlength{\abovecaptionskip}{-0.1cm}  
    \setlength{\belowcaptionskip}{-0.1cm} 
  \centering
  \caption{\rmfamily Capacity Credit and Economic Performance of ES Compared with Different SoC Withholding}
    \setlength{\tabcolsep}{0.8mm}{
    \begin{tabular}{cccccc}
    \toprule
    \multirow{2}[4]{*}{\makecell{SoC\\ Withholding}} & \multirow{2}[4]{*}{Indices} & \multicolumn{4}{c}{Power and Energy Capacity} \\
\cmidrule{3-6}          &       & 10\%-4h & 10\%-12h & 30\%-4h & 30\%-12h \\
    \midrule
    0\% & \makecell{CC (\%)\\Profit (\$$10^6$)} & \makecell{40.6\\246.1} & \makecell{91.1\\510.7} & \makecell{34.3\\572.9} & \makecell{70.1\\988.7} \\
    20\% & \makecell{CC (\%)\\Profit (\$$10^6$)} & \makecell{42.5\\256.5} & \makecell{91.3\\511.5} & \makecell{35.8\\593.3} &\makecell{70.3\\990.1} \\
    40\% & \makecell{CC (\%)\\Profit (\$$10^6$)} & \makecell{44.8\\269.1} & \makecell{92.0\\514.9} & \makecell{37.8\\618.5} & \makecell{70.9\\995.5}\\
    \bottomrule
    \end{tabular}
    } \label{socwithholding}
\end{table}%

\subsection{Impact of Structure of DDU on CC}\label{DDUstructure}

We further investigate how the distribution and parameters of DDU affect the CC of VES, as shown in Table~\ref{DDUdistribution}. As the amount of available DDU data increases, the distribution may progress from no assumption to unimodal, and eventually to symmetric \& unimodal. It is evident that with more DDU information, the CC values increase, as both the reformulation in Table~\ref{approximation} and dispatch strategies become less conservative. Furthermore, both an increase in the incentive effect ($\alpha$) and a decrease in the discomfort effect ($\beta$) improve the performance of VES, with the results being more sensitive to changes in the discomfort effect. Specially, CC is increased by 10\%-30\% with a half decrease in the discomfort effect. By contrast, CC has only increased by 5\%-13\% with a half increase in incentive effect. This suggests that procuring VES resources with lower discomfort aversion is more profitable than simply increasing capacity remuneration. Additionally,  for long-duration VES, DDU parameters have a relatively less impact on its CC. 

Moreover, the system operator can adjust the probability level $\epsilon$ in chance constraints to balance the trade-off between available capacity and the response reliability of VES. For a 4-hour VES with 30\% rated power, increasing $\epsilon$ initially raises the CC as the increased flexibility of GES dominates. When $\epsilon$ reaches 0.2, the CC peaks at 18.5\%. Beyond this point, further increases in $\epsilon$ lead to a decline in CC. The system operator should adjust the setting of $\epsilon$ based on the market practice.
\begin{table}[!ht]
      \setlength{\abovecaptionskip}{-0.1cm}  
    \setlength{\belowcaptionskip}{-0.1cm} 
  \centering
  \caption{\rmfamily Capacity Credit of VES Compared with Different Distributions and Parameters of DDU}
    \setlength{\tabcolsep}{1mm}{
    \begin{tabular}{ccccccc}
    \toprule
    \multirow{2}[4]{*}{\makecell{Distribution\\ Type}} & \multirow{2}[4]{*}{$\alpha$} & \multirow{2}[4]{*}{$\beta$} & \multicolumn{4}{c}{Power \& Energy Capacity} \\
\cmidrule{4-7}          &       &       & 10\%-4h & 30\%-4h & 10\%-12h & 30\%-12h \\
    \midrule
    \multirow{3}[2]{*}{No Assumption} & 1.0   & 4.0   & {18.8\%} & {14.7\%} & {33.7\%} & {22.9\%} \\
          & 1.0   & 2.0   & {24.0\%} & {18.5\%} & {41.0\%} & {27.0\%} \\
          & 0.5   & 2.0   & {20.6\%} & {16.4\%} & {38.9\%} & {26.1\%} \\
    \midrule
    \multirow{3}[2]{*}{Unimodal} & 1.0   & 4.0   & {21.8\%} & {16.8\%} & {37.1\%} & {24.8\%} \\
          & 1.0   & 2.0   & {28.1\%} & {21.0\%} & {44.4\%} & {28.5\%} \\
          & 0.5   & 2.0   & {24.1\%} & {18.7\%} & {42.6\%} & {27.8\%} \\
    \midrule
    \multirow{3}[2]{*}{\makecell{Symmetric\&\\Unimodal}} & 1.0   & 4.0   & {23.1\%} & {17.7\%} & {38.3\%} & {25.5\%} \\
          & 1.0   & 2.0   & {29.8\%} & {21.9\%} & {45.5\%} & {28.9\%} \\
          & 0.5   & 2.0   & {25.6\%} & {19.7\%} & {43.9\%} & {28.4\%} \\
    \bottomrule
    \end{tabular}
    }\label{DDUdistribution}
\end{table}%

\subsection{Impact of GES Locating \& Network Capacity on CC}\label{network}

The CC values of ES with different locations and transmission line capacity are summarized in Table~\ref{networkresult}. It is observed that CC declines with decreased line capacity, and FTM-ES has a higher CC than BTM-ES due to its distributed location, which reduces the impact of network congestion. Furthermore, while a relatively larger decrease in CC (around 10\%) is observed for short-duration ES, long-duration ES is also affected, though to a lesser extent.  The above analysis highlights the importance of considering network limitations for short-duration ES, and it suggests procuring more BTM-ES, rather than relying solely on FTM-ES in system planning.
\begin{table}[!ht]
      \setlength{\abovecaptionskip}{-0.1cm}  
    \setlength{\belowcaptionskip}{-0.1cm} 
  \centering
  \caption{\rmfamily Capacity Credit of ES Compared with Different Locations and Transmission Line Capacity}
    \setlength{\tabcolsep}{1mm}{
    \begin{tabular}{cccccc}
    \toprule
    \multirow{2}[4]{*}{\makecell{Line Capacity}} & \multirow{2}[4]{*}{\makecell{ES Type}} & \multicolumn{4}{c}{Power \& Energy Capacity} \\
\cmidrule{3-6}          &       & 10\%-4h & 30\%-4h & 10\%-12h & 30\%-12h \\
    \midrule
    100\% & \makecell{FTM\\BTM} & \makecell{42.5\%\\42.9\%} & \makecell{35.8\%\\35.8\%} & \makecell{91.3\%\\91.4\%} & \makecell{70.3\%\\70.3\%} \\
    70\%  & \makecell{FTM\\BTM} & \makecell{40.2\%\\41.0\%} & \makecell{33.6\%\\34.1\%} & \makecell{90.5\%\\91.0\%} & \makecell{69.8\%\\69.9\%}  \\
    50\%  & \makecell{FTM\\BTM} & \makecell{38.1\%\\39.6\%} & \makecell{31.2\%\\32.9\%} & \makecell{89.7\%\\90.6\%} & \makecell{69.1\%\\69.4\%} \\
    \bottomrule
    \end{tabular}
    }\label{networkresult}
\end{table}%

\subsection{Impact of Load Factor \& Correlation Coefficient on CC}

Given that VES is derived from the load, it is crucial to address the load factor of VES and the correlation coefficient between load and VES. We compare the CC of different VES: $\text{TCL}$-$\text{VES}^{1}$ (cooling in summer, used as the baseline in Section~\ref{case}), $\text{TCL}$-$\text{VES}^{2}$ (heating in winter), $\text{TCL}$-$\text{VES}^{3}$ (cooling in summer \& heating in winter), and $\text{TCL}$-$\text{VES}^{4}$ (heating year-round). As shown in Table~\ref{loadfactor}, for the 10\% rated power 4-hour $\text{TCL}$-$\text{VES}^{1,2}$ with the same load factor, a positive correlation coefficient results in higher CC value (40.8\%), whereas a negative correlation coefficient leads to lower CC value (21.8\%). Furthermore, as the load factor increases, the CC shows a significant rise from around 20-30\% to approximately 60-100\%, with 12-hour $\text{VES}^{3,4}$ even reaching 100\% CC. Thus, it is more economically efficient to procure VES with a higher load factor and positive correlation with the load.  This analysis also highlights the importance of temporal CC evaluation for GES.

\begin{table}[!ht]
      \setlength{\abovecaptionskip}{-0.1cm}  
    \setlength{\belowcaptionskip}{-0.1cm} 
  \centering
  \caption{\rmfamily Capacity Credit of VES Compared with Different Load Factors and Correlation Factors}
    \setlength{\tabcolsep}{0.3mm}{
    \begin{tabular}{ccccccc}
    \toprule
    \multirow{2}[4]{*}{GES Type} & \multirow{2}[4]{*}{\makecell{Load\\ Factor}} & \multirow{2}[4]{*}{\makecell{Correlation\\ Factor}} & \multicolumn{4}{c}{Power \& Energy Capacity} \\
\cmidrule{4-7}          &       &       & 10\%-4h & 30\%-4h & 10\%-12h & 30\%-12h \\
    \midrule
    $\text{TCL}$-$\text{VES}^{1}$ & 0.54  & -0.34  & 21.8\% & 16.8\% & 37.1\% & 24.8\%  \\
    $\text{TCL}$-$\text{VES}^{2}$ & 0.54  & 0.38  & 40.8\% & 33.9\% & 70.3\% & 56.4\%   \\
    $\text{TCL}$-$\text{VES}^{3}$ & 0.92  & 0.21  & 58.2\% & 50.7\% & 100\% & 90.7\%    \\
    $\text{TCL}$-$\text{VES}^{4}$ & 1.00  & 0.11  & 60.2\% & 52.6\% & 100\% & 94.8\%  \\
    \bottomrule
    \end{tabular}
    }\label{loadfactor}
\end{table}%

\subsection{Impact of Energy Losses Factors on CC}

The impact of efficiency ($\eta$) on the CC of ES and the impact of self-discharge rate ($\varepsilon$) on the CC of VES are summarized in Table~\ref{efficiency}. It is observed that ES with higher efficiency tends to have a higher CC. For short-duration ES, improvements in efficiency lead to an absolute increase of around 15\% in CC, while for long-duration ES, the increase is approximately 30\%. This highlights the greater importance of improving the efficiency of long-duration ES, such as hydrogen storage and pumped hydro storage. Meanwhile, when the energy capacity remains unchanged, variations in the self-discharge rate are akin to changes in indoor temperature ranges for TCL-VES. The results indicate that higher self-discharge rates are equivalent to wider indoor temperature ranges (increased flexibility), which leads to higher CC values. Furthermore, the impact of the self-discharge rate on CC is relatively minor for VES with higher power capacity. 
\begin{table}[!ht]
      \setlength{\abovecaptionskip}{-0.1cm}  
    \setlength{\belowcaptionskip}{-0.1cm} 
  \centering
  \caption{\rmfamily Capacity Credit of GES Compared with Different Energy Loss Factors}
    \setlength{\tabcolsep}{1.6mm}{
    \begin{tabular}{cccccc}
    \toprule
    \multirow{2}[4]{*}{GES Type} & \multirow{2}[4]{*}{$\eta$/$\varepsilon$} & \multicolumn{4}{c}{Power \& Energy Capacity} \\
\cmidrule{3-6}          &       & 10\%-4h & 30\%-4h & 10\%-12h & 30\%-12h \\
    \midrule
    \multirow{3}[1]{*}{FTM-ES} & 0.7   & 26.3\% & 22.2\% & 58.2\% & 45.6\% \\
          & 0.8   & 36.8\% & 31.1\% & 74.0\% & 57.8\% \\
          & 0.9   & 42.5\% & 35.8\% & 91.3\% & 70.3\%  \\
          \midrule
    \multirow{3}[1]{*}{$\text{TCL}$-$\text{VES}^{1}$}     & 0.4   & 21.8\% & 16.8\% & 37.1\% & 24.8\%    \\
        & 0.6   & 27.9\% & 20.8\% & 42.4\% & 26.9\%    \\
          & 0.8   & 33.3\% & 23.5\% & 44.0\% & 27.6\%   \\
    \bottomrule
    \end{tabular}
    }\label{efficiency}
\end{table}%

\subsection{Impact of Penetration of RES on CC}

\begin{figure}[!ht]
      \setlength{\abovecaptionskip}{-0.1cm}  
    \setlength{\belowcaptionskip}{-0.1cm} 
  \begin{center}
\includegraphics[width=1\columnwidth]{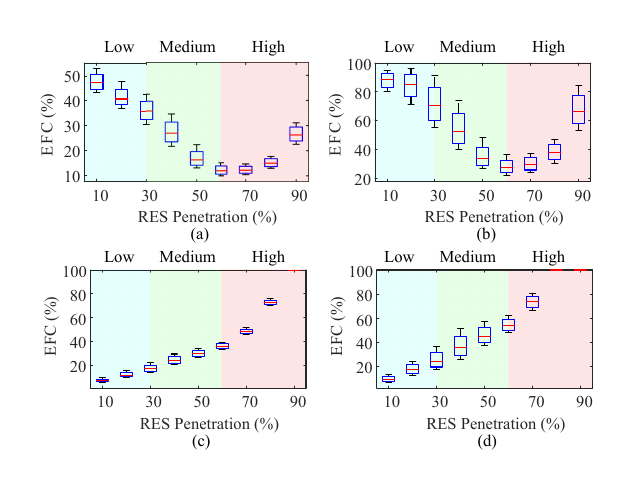}
\caption{CC of ES and VES with different RES penetration: (a) 4-h FTM-ES, (b) 12-h FTM-ES, (c) 4-h $\text{TCL}$-$\text{VES}^{1}$ and (d) 12-h $\text{TCL}$-$\text{VES}^{1}$.}\label{penetration}
  \end{center}\vspace{-0.5cm}
\end{figure}

We further investigate the CC of GES under different RES penetration while keeping the forms and composition of RES technologies (e.g., wind, solar) and load (e.g., industrial, commercial, and residential) unchanged. As shown in Fig.~\ref{penetration}, as RES penetration increases, the CC of ES experiences a slight decrease at low RES penetration levels (10\%-30\%), followed by a sharp decline at medium penetration levels (30\%-60\%), bottoming out at around 10\%-20\%. It then gradually grows at high RES penetration (70\%-100\%). This suggests that the generation substitution ability of ES decreases as RES penetration increases. However, as conventional generation becomes increasingly scarce, the CC of ES begins to rise. Furthermore, VES shows distinctly different patterns, with its CC increasing as RES penetration rises. This is because, in our case study setup, VES capacity expands with rising RES penetration, while the residual fixed load capacity decreases.  This enables a higher share of load flexibility, contributing to the increase in the CC of VES. It is also observed that long-duration GES outperforms the others and even achieve 60\%-100\% CC in the high RES penetration stage, which is a promising solution for the future decarbonized power system.

\section{Conclusion}\label{conclusion}

In this paper, a novel CC evaluation framework is proposed for GES, which fully incorporates cross-market strategic behavior and DDU of GES. Within this framework, we present a market-oriented, risk-averse coordinated dispatch method to coordinate price arbitrage with SoC withholding in the energy market and capacity provision in the capacity market. Meanwhile, we introduce chance-constrained optimization with a data-driven tractable solution methodology to mitigate risk from DDU during capacity market calls. Case studies based on the proposed practical GES performance simulation show that overlooking these factors will result in around 20\%-50\% decline in practical CC and profit. While the proposed method yields accurate CC and improves adequacy and economic performance by addressing cross-market strategic behavior and mitigating the risk from DDU. The results of the proposed ESCS indicate that VES is only 40\%-60\% equivalent to ES. 
Additionally, the CC of ES is sensitive to factors such as power and energy ratings, and DDU structure. These findings offer valuable insights for capacity market decision-making.

Future work will extend the CC evaluation framework by incorporating detailed models for the network, RES technologies, and load. Additionally, we will design an effective planning model for GES based on the proposed CC evaluation.

\begin{center}
     \textbf{Appendix}
\end{center}

\begin{appendices}

\section{Proof for data-driven reformulation of chance-constrained optimization under DDU}\label{methodproof}

We first define the mean and variance from the observed samples as \(\hat{\bm{\mu}}(\bm{x})\) and \(\hat{\bm{\Sigma}}(\bm{x})\). Let \(\mathcal{Q}_{\delta}\) represent a confidence region for the first two moments, such that \((\hat{\bm{\mu}}(\bm{x})\text{,} \hat{\bm{\Sigma}}(\bm{x})) \in \mathcal{Q}_{\delta}\), with probability at least \(1 - \nu\) for \(\nu \in (0\text{,}1)\). Then, by setting \(l := 0\) and \(\epsilon := \frac{\hat{\epsilon} - \nu}{1 - \nu}\) for \(\nu \in (0, \hat{\epsilon})\) in~\eqref{conditionalchance2} of Lemma~\ref{lemma1}, and applying the robust approximation from Table II in~\cite{qi2023chance}, we obtain:
\begin{equation}
     \bm{\mu}(\bm{x})+ \overline{F}^{-1}_{\bm{x}}(\dfrac{1-\hat{\epsilon}}{1-\nu}) \sqrt{\bm{\Sigma}(\bm{x})} \leq 0\text{,} \quad \forall (\bm{\mu}(\bm{x})\text{,} \bm{\Sigma}(\bm{x})) \in \mathcal{Q}_{\nu}
\end{equation}
Next, we define a specific confidence region around \((\hat{\bm{\mu}}(\bm{x})\text{,} \hat{\bm{\Sigma}}(\bm{x}))\) as in~\eqref{confidence region}. By applying Theorem 4 in~\cite{calafiore}, we obtain~\eqref{result1}. Subsequently, setting \(\nu := 4 \exp(-(K^{\frac{1}{p}} - 2)^2/2)\) and \(\hat{\epsilon} := \epsilon\), we derive~\eqref{ddr}. According to the law of large numbers, as \(K \rightarrow \infty\), \(\hat{\bm{\mu}}(\bm{x}) \rightarrow \bm{\mu}(\bm{x})\) and \(\hat{\bm{\Sigma}}(\bm{x}) \rightarrow \bm{\Sigma}(\bm{x})\), then we approach the original formulation~\eqref{drf}. Note that the proposed data-driven reformulation is adaptive to both the data quantity and distribution information. The more data and information available, the more detailed distribution types from Table~\ref{approximation}  can be selected, making the reformulation less conservative as the approximation values decrease further down the list.
\begin{equation}\label{confidence region}
\begin{aligned}
    \mathcal{Q}_{\nu}:=&\left\{(\bm{\mu}(\bm{x})\text{,} \bm{\Sigma}(\bm{x})):\left|\bm{\mu}(\bm{x})-\hat{\bm{\mu}}(\bm{x})\right| \leq \frac{\bm{r}(\mathbf{x})}{\sqrt{K}}(2+\sqrt{2 \ln (2 / \nu)})\text{,} \right. \\
&\left. \left|\bm{\Sigma}(\bm{x})-\hat{\bm{\Sigma}}(\bm{x})\right| \leq \frac{2(r(\mathbf{x}))^2}{\sqrt{K}}(2+\sqrt{2 \ln (4 / \nu)}) \right\}
\end{aligned}
\end{equation}
\begin{equation}\label{result1}
\begin{aligned}
&\hat{\bm{\mu}}(\bm{x})+\frac{\bm{r}(\mathbf{x})}{\sqrt{K}}(2 +\sqrt{2 \ln (2 / \nu)}) \\
& +\overline{F}^{-1}_{\bm{x}}(\dfrac{1-\hat{\epsilon}}{1-\nu}) \sqrt{\hat{\bm{\Sigma}}(\bm{x})+\frac{2(\bm{r}(\mathbf{x}))^2}{\sqrt{K}}(2+\sqrt{2 \ln (4 / \nu)})} \leq 0
\end{aligned}
\end{equation}
Hence, we finish the proof.

\begin{lemma}\label{lemma1}
Let \( \xi \) be a stochastic variable with a probability distribution in the family \(\mathcal{P}(\mu^*\text{,} \sigma^{2*}\text{,} \underline{\xi}^*\text{,} \overline{\xi}^*)\), where the mean and variance are \(\mu^*\) and \(\sigma^{2*}\), and the support is \([\underline{\xi}^*\text{,} \overline{\xi}^*]\). Let \( l \) be any scalar and \(\mathcal{Q}_\nu\) be a confidence region of the first two moments such that \(\mathcal{Q}_\nu\) contains \((\mu^*\text{,} \sigma^{2*})\) with a probability of at least \(1 - \nu\)\text{,} \(\nu \in (0\text{,} 1)\). If \eqref{conditionalchance1} is satisfied, then we have \eqref{conditionalchance2}.
\begin{equation}\label{conditionalchance1}
    \inf_{P \in \mathcal{P}(\mu\text{,} \sigma^2\text{,} \underline{\xi}^*\text{,} \overline{\xi}^*)} \mathbb{P}(\xi + l \leq 0 | \xi \sim P) \geq 1 - \epsilon \text{, } \forall (\mu\text{,} \sigma^2) \in \mathcal{Q}_\nu
\end{equation}
\begin{equation}\label{conditionalchance2}
\inf_{P \in \mathcal{P}(\mu^*\text{,} \sigma^{2*}\text{,} \underline{\xi}^*\text{,} \overline{\xi}^*)} \mathbb{P}(\xi + l \leq 0 | \xi \sim P) \geq (1 - \epsilon)(1 - \nu)
\end{equation}
\end{lemma}

\noindent \textbf{Proof:}  By leveraging conditional probability theory, it is straightforward to show: 
\begin{equation}
\begin{aligned}
   & \mathbb{P}(\xi + l \leq 0 \mid \xi \sim P \in \mathcal{P}) = \mathbb{P}\left(\left(\mu^*\text{,}  \sigma^{2 *}\right)\in\mathcal{Q}_\nu \mid \xi \sim P \in \mathcal{P}\right)\times\\
   &\mathbb{P}(\xi + l \leq 0 \mid (\xi \sim P \in \mathcal{P}) \cap((\mu^*\text{,}  \sigma^{2*})\in\mathcal{Q}_\nu))+\mathbb{P}\big(\left(\mu^*\text{,}  \sigma^{2 *}\right)\notin\mathcal{Q}_\nu\mid\\
    &  \xi \sim P \in \mathcal{P}\big)\times\mathbb{P}(\xi + l \leq 0 \mid (\xi \sim P \in \mathcal{P}) \cap (  (\mu^*\text{,} \sigma^{2*})\notin\mathcal{Q}_\nu)) \\
    &\geq \mathbb{P}(\xi + l \leq 0 \mid (\xi \sim P \in \mathcal{P}) \cap ((\mu^*\text{,}  \sigma^{2*})\in\mathcal{Q}_\nu))\times \\
    &\quad \mathbb{P}((\mu^*\text{,} \sigma^{2*})\in\mathcal{Q}_\nu \mid \xi  \sim P \in \mathcal{P}) \\
    &\geq (1 - \epsilon) \Pr((\mu^*\text{,} \sigma^{2*})\in\mathcal{Q}_\nu) \\
    &\geq (1 - \epsilon)(1 - \nu)
\end{aligned}
\end{equation}
Hence, we finish the proof.

\end{appendices}

\printcredits
\vspace{0.3cm}

\noindent \textbf{Declaration of competing interest}

The authors declare that they have no known competing financial interests or personal relationships that could have appeared to influence the work reported in this paper.
\vspace{0.3cm}

\noindent \textbf{Data availability}

The original data can be downloaded from~\cite{Elia-data}. 
\vspace{0.3cm}

\noindent \textbf{Acknowledgements}

This work was supported by the National Key
Research and Development Program of China, Research and Application of Key Technologies of Intelligent Cooperative Control of Large-Scale Energy Storage System Cluster, under Grant 2021YFB2400700 and the special funding from China Postdoctoral Science
Foundation (No.2023TQ0169). M. R. Almassalkhi graciously acknowledges funding from NSF Award ECCS-2047306.

\bibliographystyle{unsrt}

\bibliography{cas_dc}



\end{document}